%% file: main.tex
\documentclass[journal,12pt,onecolumn]{IEEEtranTCOM}

\usepackage{setspace}
\ifCLASSOPTIONonecolumn \doublespace \fi
\usepackage{graphics}
\usepackage{rotating}
\usepackage{amsfonts}
\usepackage{amsmath}
\usepackage{amsthm}
\usepackage{subfigure}
\usepackage{multirow}
\allowdisplaybreaks[1]

\renewcommand{\IEEEQED}{\IEEEQEDopen}

\usepackage{cite}

\graphicspath{{Figures/}} \DeclareGraphicsExtensions{.eps,.ps}

\title{Diversity Analysis of Bit-Interleaved Coded Multiple Beamforming with Orthogonal Frequency Division Multiplexing}

\ifCLASSOPTIONconference
\author{\IEEEauthorblockN{Boyu Li and Ender Ayanoglu}\\
\IEEEauthorblockA{Center for Pervasive Communications and Computing\\
Department of Electrical Engineering and Computer Science\\
University of California, Irvine\\
Email: boyul@uci.edu, ayanoglu@uci.edu}} \else
\author{Boyu~Li,~\IEEEmembership{Member,~IEEE,}~and~Ender~Ayanoglu,~\IEEEmembership{Fellow,~IEEE}
\thanks{B. Li and E. Ayanoglu are with the Center for Pervasive Communications and Computing, Department of Electrical Engineering and Computer Science, Henry Samueli School of Engineering, University of California, Irvine, CA 92697-3975 USA (e-mail: boyul@uci.edu; ayanoglu@uci.edu).}} \fi

\begin{document}
\maketitle


\input{Abstract}



\input{Introduction}
\input{System}
\input{Diversity}

\input{Correlation}
\input{Grouping}
\input{Results}
\input{Conclusions}
\input{Acknowledgment}
\input{Appendix}


\bibliographystyle{IEEEtran}
\bibliography{IEEEabrv,Mybib}

\end{document}

%% file: Abstract.tex
\begin{abstract}

For broadband wireless communication systems, Orthogonal Frequency Division Multiplexing (OFDM) has been combined with Multi-Input Multi-Output (MIMO) techniques. Bit-Interleaved Coded Multiple Beamforming (BICMB) can achieve both spatial diversity and spatial multiplexing for flat fading MIMO channels. For frequency selective fading MIMO channels, BICMB with OFDM (BICMB-OFDM) can be applied to achieve both spatial diversity and multipath diversity, making it an important technique. However, analyzing the diversity of BICMB-OFDM is a challenging problem. In this paper, the diversity analysis of BICMB-OFDM is carried out. First, the maximum achievable diversity is derived and a full diversity condition $R_cSL \leq 1$ is proved, where $R_c$, $S$, and $L$ are the code rate, the number of parallel steams transmitted at each subcarrier, and the number of channel taps, respectively. Then, the performance degradation due to the correlation among subcarriers is investigated. Finally, the subcarrier grouping technique is employed to combat the performance degradation and provide multi-user compatibility. 

\end{abstract}

%% file: Introduction.tex
\section{Introduction} \label{sec:Introduction}

Substantial research and development interests have been drawn on Multiple-Input Multiple-Output (MIMO) systems because they can provide high spectral efficiency and performance in a given bandwidth. In a MIMO system, beamforming techniques exploiting Singular Value Decomposition (SVD) can be employed to achieve spatial multiplexing\footnotemark \footnotetext{In this paper, the term ``spatial multiplexing" is used to describe the number of spatial subchannels, as in \cite{Paulraj_ST}. Note that the term is different from ``spatial multiplexing gain" defined in \cite{Zheng_DM}.} and thereby increase the data rate, or to enhance performance, when the Channel State Information (CSI) is available at both the transmitter and receiver \cite{Jafarkhani_STC}. 

For flat fading MIMO channels, single beamforming carrying only one symbol at a time achieves full diversity \cite{Sengul_DA_SMB,Ordoez_HSAP}. However, spatial multiplexing without channel coding results in the loss of the full diversity order. To overcome the performance degradation, Bit-Interleaved Coded Multiple Beamforming (BICMB) was proposed \cite{Akay_FSPFD, Akay_BICMB, Akay_On_BICMB}. BICMB systems studied so far employ convolutional codes \cite{Lin_ECC} as channel coding, and interleave the coded bit codewords through the multiple subchannels with different diversity orders. BICMB can achieve the full diversity order as long as the code rate $R_c$ and the number of employed subchannels $S$ satisfy the condition $R_cS \leq 1$ \cite{Gresset_PBICM_MIMO, Park_DA_BICMB, Park_DA_BICMB_J}. 

If the channel is in frequency selective fading, Orthogonal Frequency Division Multiplexing (OFDM) can be used to combat the Inter-Symbol Interference (ISI) caused by multipath propagation \cite{Barry_DC}. The advantages of OFDM are well-known. In particular, 
multipath diversity can be achieved by adding channel coding \cite{Akay_BICM_OFDM, Akay_BICM_OFDM_STBC}. 
MIMO techniques have been incorporated with OFDM for all broadband wireless communication standards, i.e., the Institute of Electrical and Electronics Engineers (IEEE) 802.11 Wireless Fidelity (WiFi) standard \cite{IEEE_802_11}, the IEEE 802.16 Worldwide Interoperability for Microwave Access (WiMAX) standard \cite{IEEE_802_16}, and the Third Generation Partnership Project (3GPP) Long Term Evolution (LTE) standard \cite{3GPP_TS_36.201}. Beamforming can be combined with OFDM for frequency selective MIMO channels to combat ISI and achieve spatial diversity \cite{Zamiri_MIMO_OFDM}. Moreover, both spatial diversity and multipath diversity can be achieved by adding channel coding, e.g., BICMB with OFDM (BICMB-OFDM), \cite{Akay_BICMB_OFDM, Akay_BICMB_OFDM_CSI, Akay_BICMB}. 
Therefore, BICMB-OFDM can be an important technique for broadband wireless communication. However, the diversity analysis of BICMB-OFDM is a difficult challenge. 

In \cite{Akay_BICMB}, an initial attempt to investigate the diversity of BICMB-OFDM was based on two over-optimistic assumptions. First, the $s$th singular values realized by SVD at all subcarriers were assumed to be independent and identically distributed, which is not practical in general. Second, the bit interleaver was assumed to satisfy the condition that at least one error bit of each error event is carried on the subchannels with each index $s$, which is not always valid. Moreover, the relation between the diversity and the combination of the bit interleaver and the convolutional code was not investigated well enough as in \cite{Park_DA_BICMB, Park_DA_BICMB_J} for BICMB in flat fading MIMO channels. Unfortunately, the analysis in \cite{Park_DA_BICMB, Park_DA_BICMB_J} cannot be generalized in a straightforward manner to BICMB-OFDM for frequency selective fading MIMO channels because the diversity now jointly depends on all subcarriers. 

In this paper, the diversity analysis of BICMB-OFDM is carried out with more reasonable assumptions than \cite{Akay_BICMB}, and the relation between the diversity and the combination of the bit interleaver and the convolutional code is better investigated as in \cite{Park_DA_BICMB, Park_DA_BICMB_J}. 
First, the maximum achievable diversity is derived and the important $\alpha$-spectra directly determining the diversity are introduced. Based on the analysis, a sufficient and necessary full diversity condition, $R_cSL \leq 1$ where $S$ is the number of streams transmitted at each subcarrier and $L$ is the number of channel taps, is proved. Then, the performance degradation caused by the correlation among subcarriers is investigated. To overcome the performance degradation, the subcarrier grouping technique \cite{Wang_WMC, Goeckel_OFDM, Liu_LCP_OFDM} which also provides multi-user compatibility, is employed. 


%% file: System.tex
\section{System Model} \label{sec:system}

A BICMB system employing $N_t$ transmit and $N_r$ receive antennas, a convolutional code of code rate $R_c$, and transmitting $S$ parallel data streams has the maximum diversity order of
\begin{align}
D = (N_r - \lceil R_cS \rceil + 1)(N_t - \lceil R_cS \rceil + 1).
\label{eq:diversity_bicmb}
\end{align}
For a system description of BICMB, as well as the derivation of (\ref{eq:diversity_bicmb}), we refer the reader to \cite{Park_DA_BICMB}, \cite{Park_DA_BICMB_J}.

\ifCLASSOPTIONonecolumn
\begin{figure}[!t]
\centering \includegraphics[width = 1.0\linewidth]{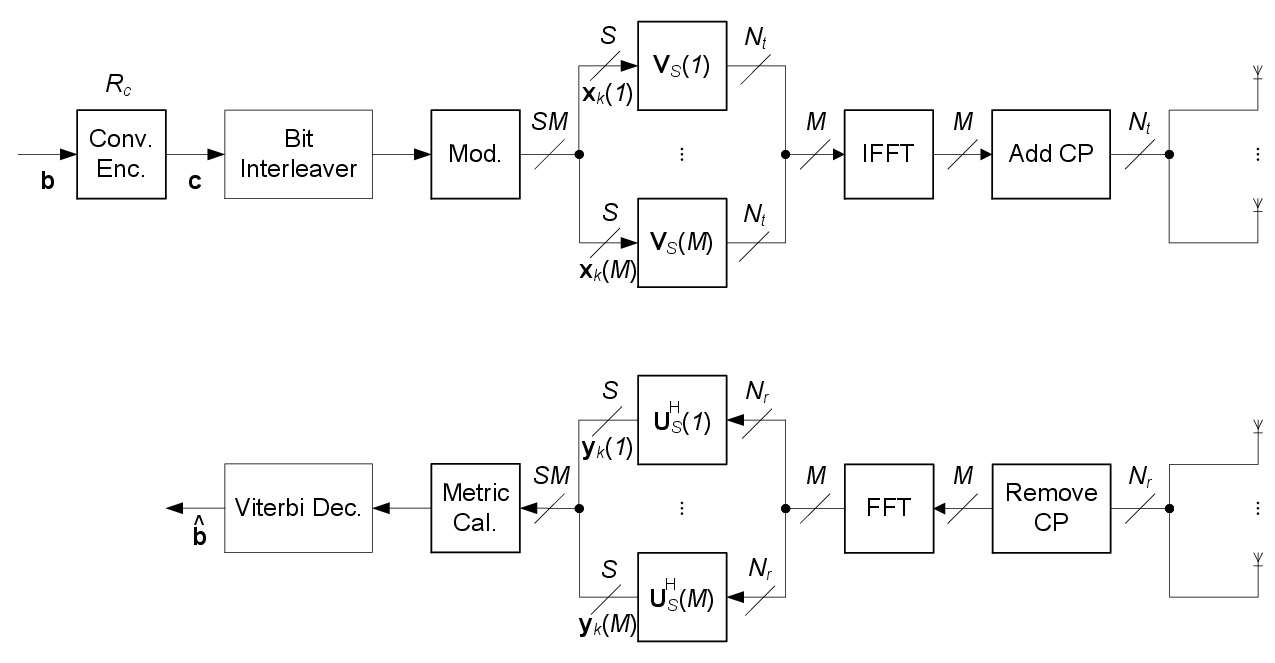}
\caption{Structure of BICMB-OFDM.} \label{fig:bicmb_ofdm}
\end{figure}
\else
\begin{figure}[!t]
\centering \includegraphics[width = 1.0\linewidth]{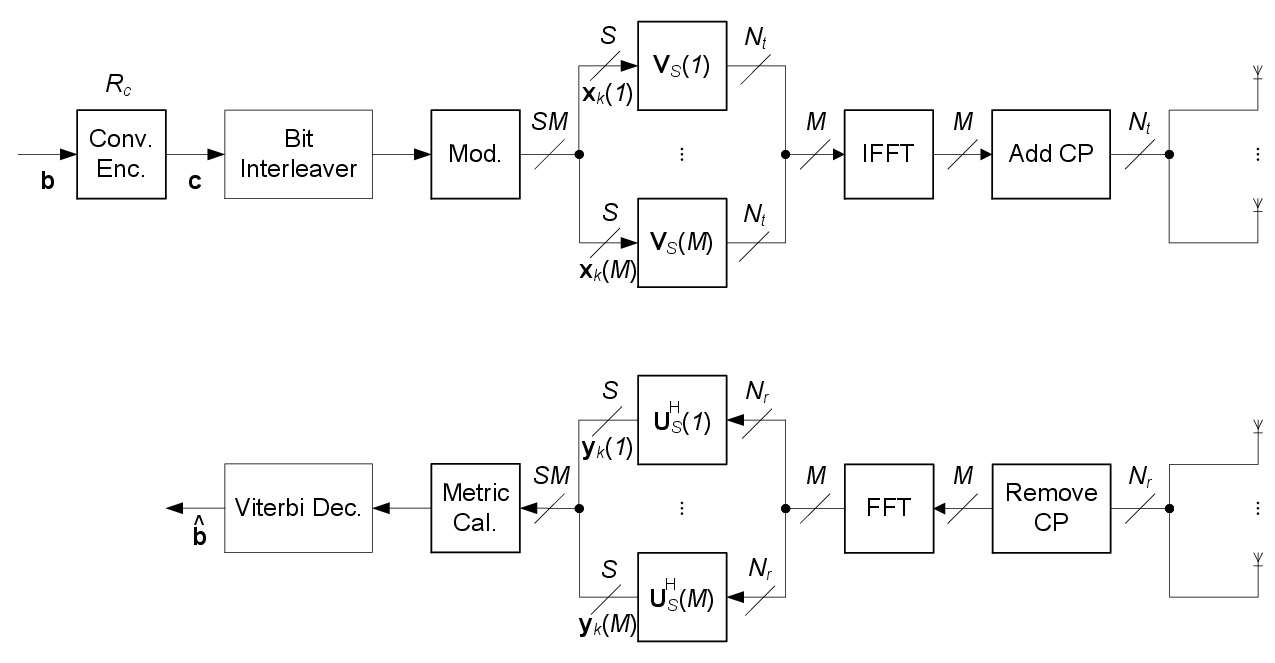}
\caption{Structure of BICMB-OFDM.} \label{fig:bicmb_ofdm}
\end{figure}
\fi

BICMB-OFDM was proposed to achieve both spatial diversity and multipath diversity for MIMO frequency selective channels, \cite{Akay_BICMB_OFDM, Akay_BICMB_OFDM_CSI, Akay_BICMB}. The structure of BICMB-OFDM is presented in Fig. \ref{fig:bicmb_ofdm}. First, the convolutional encoder of code rate $R_c$, possibly combined with a perforation matrix for a high rate punctured code \cite{Haccoun_PCC}, generates the bit codeword $\mathbf{c}$ from the information bits. Then, an interleaved bit sequence is generated by a random bit interleaver before being modulated, e.g., Quadrature Amplitude Modulation (QAM), to a symbol sequence. Assume that the symbol sequence is transmitted through $M$ subcarriers, and $S \leq \min\{N_t,N_r\}$ streams are transmitted for each subcarrier at the same time. Hence, an $S \times 1$ symbol vector $\mathbf{x}_k(m)$ is transmitted through the $m$th subcarrier at the $k$th time instant with $m = 1,\ldots,M$. The length of Cyclic Prefix (CP), which is employed by OFDM to combat ISI caused by multipath
propagation, is assumed to be $L_{cp}$ where $L_{cp} \geq L$ with $L$ denoting the number of channel taps.

The frequency selective fading MIMO channel with $L$ taps is assumed to be quasi-static Rayleigh and known by both the transmitter and the receiver, which is given by $\breve{\mathbf{H}}(l) \in \mathbb{C}^{N_r \times N_t}$ with $l=1,\ldots,L$, where $\mathbb{C}$ stands for the set of complex numbers. Let 
\begin{align}
\mathbf{H}(m)=\sum_{l=1}^{L} \breve{\mathbf{H}}(l)\exp\left({-i{2\pi (m-1) \tau_l \over MT}}\right) \label{eq:channel_frequency}
\end{align}
denote the quasi-static flat fading MIMO channel observed at the $m$th subcarrier, where $T$ denotes the sampling period, $\tau_l$ indicates the $l$th tap delay, and $i=\sqrt{-1}$ \cite{Lee_ST_BICM_OFDM}. Then, SVD beamforming is carried out for each subcarrier. The beamforming matrices at the $m$th subcarrier are determined by SVD of $\mathbf{H}(m)$, i.e., $\mathbf{H}(m) = \mathbf{U}(m) \mathbf{\Lambda}(m) \mathbf{V}^H(m)$, where the $N_r \times N_r$ matrix $\mathbf{U}(m)$ and the $N_t \times N_t$ matrix $\mathbf{V}(m)$ are unitary, and the $N_r \times N_t$ matrix $\mathbf{\Lambda}(m)$ is diagonal rectangular whose $s$th diagonal element, $\lambda_s(m) \in \mathbb{R}^+$, is a singular value of $\mathbf{H}(m)$ or a square root of the eigenvalue $\phi_s(m)$ of $\mathbf{H}(m)\mathbf{H}^H(m)$ in decreasing order with $s=1,\ldots,S$, where $\mathbb{R}^+$ denotes the set of positive real numbers. When $S$ streams are transmitted at the same time, the first $S$ columns of $\mathbf{U}(m)$ and $\mathbf{V}(m)$, i.e., $\mathbf{U}_S(m)$ and $\mathbf{V}_S(m)$, are chosen as beamforming matrices at the receiver and transmitter at the $m$th subcarrier, respectively.

For each subcarrier, the multiplications with beamforming matrices are carried out before executing the Inverse Fast Fourier Transform (IFFT) and adding CP at the transmitter, and after executing the Fast Fourier Transform (FFT) and removing CP at the receiver, respectively. Therefore, the system input-output relation for the $m$th subcarrier at the $k$th time instant is
\begin{align}
y_{k,s}(m) = {\lambda}_s(m) x_{k,s}(m) + n_{k,s}(m), \label{eq:detected_symbol}
\end{align}
with $s=1,\ldots,S$, where $y_{k,s}(m)$ and $x_{k,s}(m)$ are the $s$th element of the $S \times 1$ received symbol vector $\mathbf{y}_k(m)$ and the transmitted symbol vector $\mathbf{x}_k(m)$ respectively, and $n_{k,s}(m)$ is the additive white Gaussian noise with zero mean and variance $N_0=N_t/\gamma$ \cite{Liu_STF_OFDM}, with $\gamma$ denoting the received Signal-to-Noise Ratio (SNR) over all the receive antennas. Note that the total transmitted power is scaled by $N_t$ in order to make the received SNR $\gamma$.

The location of the coded bit $c_{k'}$ within the transmitted symbol is denoted as $k' \rightarrow (k, m, s, j)$, which means that $c_{k'}$ is mapped onto the $j$th bit position on the label of $x_{k,s}(m)$. Let $\chi$ denote the signal set of the modulation scheme, and let $\chi_b^j$ denote a subset of $\chi$ whose labels have $b \in \{0, 1\}$ at the $j$th bit position. By using the location information and the input-output relation in (\ref{eq:detected_symbol}), the receiver calculates the Maximum Likelihood (ML) bit metrics for $c_{k'}=b \in \{0, 1\}$ as
\begin{align}
\Delta(y_{k,s}(m), c_{k'}) = \min_{x \in \chi_{c_{k'}}^j} \left| y_{k,s}(m) - {\lambda}_s(m)x \right|^2. \label{eq:ml_bit_metrics}
\end{align}

Finally, the ML decoder, which applies the soft-input Viterbi decoding \cite{Lin_ECC} to find a codeword $\mathbf{\hat{c}}$ with the minimum sum weight and its corresponding information bit sequence $\mathbf{\hat{b}}$, uses the bit metrics calculated by  (\ref{eq:ml_bit_metrics}) and makes decisions according to the rule given by \cite{Caire_BICM} as
\begin{align}
\mathbf{\hat{c}} = \arg\min_{\mathbf{c}} \sum_{k'} \Delta(y_{k,s}(m), c_{k'}).
\label{eq:decision_rule}
\end{align}

%% file: Diversity.tex
\section{Maximum Achievable Diversity of BICMB-OFDM} \label{sec:diversity} 

The performance of BICMB-OFDM is bounded by the union of the Pairwise Error Probability (PEP) corresponding to each error event \cite{Akay_BICMB_OFDM, Akay_BICMB_OFDM_CSI, Akay_BICMB}. In particular, the overall diversity order is dominated by the pairwise errors which have the smallest negative exponent of SNR in their PEP representations. 
Define an $M \times S$ matrix $\mathbf{A}$, whose element $\alpha_{m,s}$ denotes the number of distinct bits transmitting through the $s$th subchannel of the $m$th subcarrier for an error path, which implies that $\sum_{m=1}^M \sum_{s=1}^S \alpha_{m,s} = d_H$. Let $\mathbf{a}^T_m$ denote the $m$th row of $\mathbf{A}$. Note that the $\alpha$-spectrum here is similar to BICMB in the case of flat fading MIMO channels introduced in \cite{Park_DA_BICMB, Park_DA_BICMB_J}. 
%
%
Consider the case that different MIMO delay spread channels are uncorrelated and have equal power, and each element of each tap is statistically independent and modeled as a complex Gaussian random variable with zero mean and variance $1/L$, 
then all subcarriers are independent in the case of $L=M$ \cite{Li_DA_BICMB_OFDM, Li_DA_BICMB_OFDM_arXiv}. In the following part of this section, this special case is considered. Although this special case is not practical, its diversity analysis provides the maximum achievable diversity for the practical case. The reason is that correlation among subcarriers for the practical case has a negative effect on performance, which will be discussed in Section \ref{sec:correlation}.

In the case of $L=M$, an upper bound of PEP is
\begin{align}
\mathrm{Pr} \left( \mathbf{c} \rightarrow \hat{\mathbf{c}} \right) & \leq \prod_{m,\mathbf{a}_m \neq \mathbf{0}} {\zeta}_m \left( \frac{ d^2_{min} \alpha_{m,min}} {4 N_t} \gamma \right)^{-D_m},
\label{eq:pep_diversity}
\end{align}
with $D_m=(N_r-\delta_m+1)(N_t-\delta_m+1)$, where $d_{min}$ is the minimum Euclidean distance \cite{Proakis_DC} in the constellation, $\alpha_{m,min}$ denotes the minimum non-zero element in $\mathbf{a}_m$, $\delta_m$ denotes the index of the first non-zero element in $\mathbf{a}_m$, and ${\zeta}_m$ is a constant \cite{Li_DA_BICMB_OFDM, Li_DA_BICMB_OFDM_arXiv}. Therefore, the diversity can be easily found from (\ref{eq:pep_diversity}), which is 
\begin{align}
D = \sum_{m,\mathbf{a}_m \neq \mathbf{0}} D_m.
\label{eq:diversity}
\end{align}

Because the error paths with the worst diversity order dominate the performance, the results of (\ref{eq:pep_diversity}) and (\ref{eq:diversity}) show that the maximum achievable diversity of BICMB-OFDM is directly decided by the $\alpha$-spectra. Note that the $\alpha$-spectra are related with the bit interleaver and the trellis structure of the convolutional code, and they can be derived by a similar approach to BICMB in the case of flat fading MIMO channels presented in \cite{Park_DA_BICMB}, or by computer search. An example is provided below to show the relation between the $\alpha$-spectra and the diversity. 
\\\textbf{Example:} Consider the parameters $N_t=N_r=S=L=M=2$. Assume that the $R_c=1/2$ convolutional code with generator polynomial $(5,7)$ in octal is employed, and the bit interleaver applies simple bit rotation, i.e., the $s$th bit in the $m$th group of $S$ bits are transmitted through the $s$th subchannels at the $m$th subcarrier for one time instant. In this case, the dominant error path has the $\alpha$-spectrum $\mathbf{A}=[0 \, 1; \, 2 \, 2]$, which implies that $\delta_1=2$ and $\delta_2=1$. Hence, $D_1=1$ and $D_2=4$ in (\ref{eq:pep_diversity}). Therefore, the maximum achievable diversity order in (\ref{eq:diversity}) is $D=D_1+D_2=5$.   

\subsection{The $\alpha$-spectra} \label{subsec:alpha}

\ifCLASSOPTIONonecolumn
\begin{figure}[!t]
\centering \includegraphics[width = 0.5\linewidth]{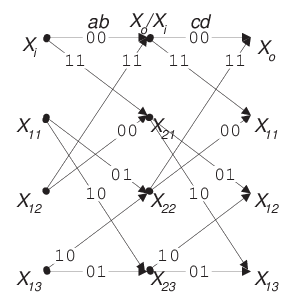} 
\caption{Trellis of $4$-state $R_c=1/2$ convolutional code with $4$ streams}
\label{fig:trellis_alpha}
\end{figure}
\else
\begin{figure}[!t]
\centering \includegraphics[width = 0.6\linewidth]{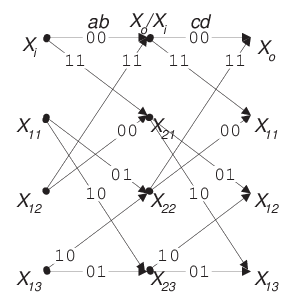} 
\caption{Trellis of $4$-state $R_c=1/2$ convolutional code with $4$ streams}
\label{fig:trellis_alpha}
\end{figure}
\fi

A method to derive the $\alpha$-spectra is illustrated by the following simple example. 
\\\textbf{Example:} Consider that the system is composed of a $4$-state $R_c=1/2$ convolutional encoder and a spatial de-multiplexer rotating with an order of $a$, $b$, $c$, and $d$ which represent the four streams of transmission. Fig. \ref{fig:trellis_alpha} represents a trellis diagram of this convolutional encoder for one period at the steady state. Since a convolutional code is linear, the all-zeros codeword is assumed to be the input to the encoder. To find a transfer function of a convolutional code and a spatial de-multiplexer, the branches are labeled as a combination of $a^{\beta_a}$, $b^{\beta_b}$, $c^{\beta_c}$, and $d^{\beta_d}$, where the exponent denotes the number of usage for each subchannel which causes error decoding. Additionally, $Z^{\beta_Z}$, whose exponent satisfies $\beta_Z = \beta_a+\beta_b+\beta_c+\beta_d$, is included to get the relation between the Hamming distance $d_H$ \cite{Lin_ECC} of two codewords and $\alpha$-spectrum of an error event. Furthermore, the non-zero states are symbolically labeled from $X_{11}$ to $X_{23}$ as in Fig. \ref{fig:trellis_alpha}, while the zero state is labeled as $X_i$ if branches split and $X_o$ if branches merge, also as shown in Fig. \ref{fig:trellis_alpha}.

Define $\mathbf{x}$ = $ \left[X_{11} ,\, X_{12} ,\, X_{13} ,\ X_{21} ,\ X_{22} ,\ X_{23}\right]^T$. Then, one state equation is given by the matrix equation
\ifCLASSOPTIONonecolumn
\begin{align}
\mathbf{x} &= \mathbf{Fx} + \mathbf{t}X_i = \left[
\begin{array}{cccccc}
0 & 0 & 0 & 0 & 1 & 0 \\
0 & 0 & 0 & dZ & 0 & cZ \\
0 & 0 & 0 & cZ & 0 & dZ \\
0 & 1 & 0 & 0 & 0 & 0 \\
bZ & 0 & aZ & 0 & 0 & 0 \\
aZ & 0 & bZ & 0 & 0 & 0 \end{array} \right] \mathbf{x} + \left[
\begin{array}{c}
cdZ^2 \\ 0 \\ 0 \\ abZ^2 \\ 0 \\ 0\end{array} \right] X_{i}.
\label{eq:transfer_i}
\end{align}
\else
\begin{align}
\mathbf{x} &= \mathbf{Fx} + \mathbf{t}X_i \nonumber \\
&= \left[
\begin{array}{cccccc}
0 & 0 & 0 & 0 & 1 & 0 \\
0 & 0 & 0 & dZ & 0 & cZ \\
0 & 0 & 0 & cZ & 0 & dZ \\
0 & 1 & 0 & 0 & 0 & 0 \\
bZ & 0 & aZ & 0 & 0 & 0 \\
aZ & 0 & bZ & 0 & 0 & 0 \end{array} \right] \mathbf{x} + \left[
\begin{array}{c}
cdZ^2 \\ 0 \\ 0 \\ abZ^2 \\ 0 \\ 0\end{array} \right] X_{i}.
\label{eq:transfer_i}
\end{align}
\fi
Similarly,
\begin{align}
X_{o} = \mathbf{gx} = \left[ \begin{array}{cccccc} 0 & abZ^2 & 0 & 0
& cdZ^2 & 0 \end{array} \right] \mathbf{x}. 
\label{eq:transfer_o}
\end{align}
The transfer function is represented in closed form by using the method in \cite{Haccoun_PCC} as
\ifCLASSOPTIONonecolumn
\begin{align}
\mathbf{T}(a,b,c,d,Z) &= \mathbf{g} \left[\mathbf{I}-\mathbf{F}\right]^{-1} \mathbf{t} \nonumber\\
&=\mathbf{gt} + \sum\limits_{u=1}^{\infty} \mathbf{gF}^{u} \mathbf{t} \nonumber\\
&= Z^5 (a^2 b^2 d + b c^2 d^2) + \nonumber\\
&\quad\,\, Z^6 (2 a^2 b c^2 d + a^2 b^2 d^2 + b^2 c^2 d^2) + \nonumber\\
&\quad\,\, Z^7 (a^2 b^3 c^2 + 2 a^2 b^2 c^2 d + 2 a^2 b c^2 d^2 + b^3 c^2 d^2 + a^2 b^2 d^3 + a^2 c^2 d^3) + \nonumber\\
&\quad\,\, Z^8 (a^4 b^2 c^2 + 4 a^2 b^3 c^2 d + 4 a^2 b^2 c^2 d^2 + b^4 c^2 d^2 + a^2 c^4 d^2 + 4 a^2 b c^2 d^3 + a^2 b^2 d^4) + \cdots ,
\label{eq:transfer_example}
\end{align}
\else
\begin{align}
\mathbf{T}&(a,b,c,d,Z) = \mathbf{g} \left[\mathbf{I}-\mathbf{F}\right]^{-1} \mathbf{t} \nonumber\\
&=\mathbf{gt} + \sum\limits_{u=1}^{\infty} \mathbf{gF}^{u} \mathbf{t} \nonumber\\
&= Z^5 (a^2 b^2 d + b c^2 d^2) + \nonumber\\
&\quad\,\, Z^6 (2 a^2 b c^2 d + a^2 b^2 d^2 + b^2 c^2 d^2) +\nonumber\\
&\quad\,\, Z^7 (a^2 b^3 c^2 + 2 a^2 b^2 c^2 d + 2 a^2 b c^2 d^2 + \nonumber\\
&\qquad b^3 c^2 d^2 + a^2 b^2 d^3 + a^2 c^2 d^3) + \nonumber\\
&\quad\,\, Z^8 (a^4 b^2 c^2 + 4 a^2 b^3 c^2 d + 4 a^2 b^2 c^2 d^2 + \nonumber\\
&\qquad b^4 c^2 d^2 + a^2 c^4 d^2 + 4 a^2 b c^2 d^3 + a^2 b^2 d^4) + \cdots ,
\label{eq:transfer_example}
\end{align}
\fi
where $[\mathbf{I}-\mathbf{F}]^{-1}$ can be expanded as $\mathbf{I} + \mathbf{F} + \mathbf{F}^2 + \cdots$ through an infinite power series of matrices. 

Consider BICMB-OFDM with parameters $N_t=N_r=S=L=M=2$. Therefore, four streams are transmitted simultaneously. Assume that $a$ and $b$ are assigned to be the first and second streams transmitted by the first subcarrier respectively, and $c$ and $d$ are assigned to be the first and second streams transmitted by the second subcarrier respectively. The $\alpha$-spectra can be figured out from the transfer function. For example, there are two error events with $d_h=5$ whose $\alpha$-spectra are $[2\, 2; \, 0\, 1]$ and $[0\, 1; \, 2\, 2]$ respectively.

This method can be applied to any $\mathcal{K}$-state $R_c=k_c / n_c$ convolutional code where $k_c$ and $n_c$ are positive integers with $k_c < n_c$ and BICMB-OFDM with $M$ subcarriers and $S$ streams transmitted at each subcarrier. Note that $R_c = k_c/n_c$ implies that each $k_c$ sections in the trellis of the convolutional code generates $n_c$ coded bits. If the spatial de-multiplexer is not a random switch for the whole packet, the period of the spatial de-multiplexer is an integer multiple of the Least Common Multiple (LCM) of $n_c$ and $SM$. Note that a period of the interleaver is restricted to an integer multiple of the trellis sections. Define $Q$ = $\mathrm{LCM}(n_c, SM)$ as the number of coded bits for a minimum period. Then, the dimension of the vector $\mathbf{x}$ is $nQ(\mathcal{K}-1)k_c/n_c$ where $n$ is a positive integer.

\subsection{Full Diversity Condition} \label{subsec:full_diversity}

Note that since the subcarriers are independent of each other in the case of $M=L$, the MIMO channels in (\ref{eq:channel_frequency}) in the frequency domain can be addressed as the block fading MIMO channels considered in \cite{Gresset_STC_BICM}. According to \cite{Gresset_STC_BICM}, the full diversity is $N_rN_tL$, which is consistent with the full diversity of frequency selective fading MIMO channels \cite{Jafarkhani_STC}. Based on the results of (\ref{eq:pep_diversity}) and (\ref{eq:diversity}), full diversity of $N_rN_tL$ can be achieved by BICMB-OFDM if and only if all entries in the first column of the $\mathbf{A}$ matrix are non-zero, i.e., $\alpha_{m,1} \neq 0, \forall m$, for all error events. To meet such requirements of the $\alpha$-spectra, the condition $R_cSL \leq 1$ needs to be satisfied. In the following, the proof of the full diversity condition with the rate of the convolutional code is provided. 

\begin{IEEEproof}
To prove the necessity, assume that an information bit sequence $\mathbf{b}$ with length $N_b=JR_cSL$ is transmitted, then a bit sequence $\mathbf{c}_{m,s}$ containing $J$ bits is transmitted at the $s$th subchannel of the $m$th subcarrier. If $R_cSL>1$, because the number of different codewords $2^{N_b}$ is larger than the number of different bit sequences $\mathbf{c}_{m,s}$, $2^J$, there always exists at least a pair of codewords which results in the same $\mathbf{c}_{m,s}$. As a result, the pairs of codewords with the same $\mathbf{c}_{m,1}$ result in $\alpha_{m,1} = 0$, and therefore cause full diversity loss.  

To prove the sufficiency, consider a bit interleaver employing simple rotation with the condition $R_cSL \leq 1$. Simple rotation means that
the coded bits are multiplexed for each subchannel at each subcarrier, with increasing order of subchannels first and then subcarriers, i.e., the first subchannel of the first subcarrier, $\cdots$, the last subchannel of the first subcarrier, the first subchannel of the second subcarrier,  and so on. Because $R_c \leq 1/(SL)$, the number of coded bits generated from each section in the trellis structure of the convolutional code is no less than $SL$. In this case, all subchannels at each subcarrier could be assigned to one section in the trellis structure of the convolutional code. Since the trellis of the convolutional code can be designed such that the coded bits generated from the first branch splitting from the zero state are all errored bits of an error event, each subchannel of all subcarriers could be used at least once, which guarantees $\alpha_{m,1} \neq 0, \forall m$, for all error paths. Therefore, full diversity can be achieved.

This concludes the proof.
\end{IEEEproof}

The proof of the necessity above implies that in the case of $R_cSL > 1$, there always exists at least an error path with no errored bits transmitted through the first subchannel of a subcarrier. Therefore, full diversity cannot be achieved. In this case, the bit interleaver should be designed such that consecutive coded bits are transmitted over different subchannels of different subcarriers to provide the maximum achievable diversity, which depends on the $\alpha$-spectra.

To better illustrate the proof of the sufficiency above, a simple example is given below.
\\\textbf{Example:} Consider the parameters $N_t=N_r=L=M=2$ and $S=1$. Also assume that the $R_c=1/2$ convolutional code with generator polynomial $(5,7)$ in octal is used. Note that the trellis structure of this code can be represented by one section of the trellis in Fig. \ref{fig:trellis_alpha}. Since $R_c = 1/2  \leq 1/(SL)$, both of the $SL=2$ subchannels could be assigned to one section in the trellis structure of the convolutional code. Assume that $a$ and $b$ are assigned to be the streams transmitted by the first and the second subcarriers respectively. Then, the trellis diagram for one period at the steady state of this combination of the convolutional code and the bit interleaver can actually be represented as the first section in Fig. \ref{fig:trellis_alpha}. Now, due to the fact that the coded bits generated from the first branch splitting from the zero state are all errored bits of an error event, the full diversity requirements $\alpha_{m,1} \neq 0, \forall m$ for all error paths are satisfied. 

Note that the full diversity condition of BICMB for flat fading MIMO channels is $R_cS \leq 1$ \cite{Gresset_PBICM_MIMO, Park_DA_BICMB, Park_DA_BICMB_J}. Now the condition $R_cSL \leq 1$ of BICMB-OFDM for frequency selective fading MIMO channels involves the number of channel taps $L$. It is not a simple generalization because the total parallel steams in actually $SM$ instead of $SL$. Moreover, a similar full diversity condition $R_cN_tL \leq 1$ for block fading MIMO channels was derived in \cite{Gresset_STC_BICM}. Note that the condition $R_cSL \leq 1$ of BICMB-OFDM for frequency selective fading MIMO channels is tighter since $S\leq\min\{N_t, N_r\}$.

%% file: Correlation.tex
\section{Negative Effect of Subcarrier Correlation} \label{sec:correlation}

In practice, $M$ is always much larger than $L$. In this case, correlation exists among subcarriers \cite{Edfors_OFDM_SVD, Li_DA_BICMB_OFDM, Li_DA_BICMB_OFDM_arXiv}. Hence, to calculate PEP, the joint Probability Density Function (PDF) of diagonal elements in $\mathbf{\Lambda}(m) \mathbf{\Lambda}^H(m)$ for all $m$ satisfying $\mathbf{a}_m \neq \mathbf{0}$, which are eigenvalues of a set of correlated Wishart matrices \cite{Edelman_Eigen}, is required \cite{Li_DA_BICMB_OFDM, Li_DA_BICMB_OFDM_arXiv}. However, this is an extremely difficult problem. The joint PDF of two correlated Wishart matrices is given in \cite{Smith_Wishart, Kuo_Wishart, Chiani_MIMO}, which is already highly complicated. To the best of our knowledge, the joint PDF of more than two correlated Wishart matrices is not available in the literature. The maximum diversity of an OFDM-MIMO system is known to be $N_rN_tL$ \cite{Jafarkhani_STC}. In the case of BICMB-OFDM with $M>L$, a performance degradation caused by subcarrier correlation is to be expected. Because, otherwise, the diversity can exceed the full diversity of $N_rN_tL$, which is a contradiction. In this section, the negative effect of correlation on the performance between two subcarriers is investigated to provide an intuitive insight.

Consider an error path whose $d_H$ distinct bits between two bit codewords are all transmitted through two correlated subcarriers with correlation $\rho$ in absolute value, which could be the practical case. Define $X = \max(N_t,N_r)$ and $Y = \min(N_t,N_r)$. Let $\mathbf{\Phi} = [\phi_1, \ldots, \phi_Y]$ and $\tilde{\mathbf{\Phi}} = [\tilde{\phi}_1, \ldots, \tilde{\phi}_Y]$ denote the ordered eigenvalues of the two correlated Wishart matrices $\mathbf{H}\mathbf{H}^H$ and $\tilde{\mathbf{H}}\tilde{\mathbf{H}}^H$, respectively. Note that $\phi_u = \lambda_u^2$
. Let $\mathbf{a} = [\alpha_1, \ldots, \alpha_Y]$ and $\tilde{\mathbf{a}} = [\tilde{\alpha}_1, \ldots, \tilde{\alpha}_Y]$ denote the $\alpha$-spectra of $\mathbf{\Phi}$ and $\tilde{\mathbf{\Phi}}$ respectively. Define $\mathbf{p} = [p_1, \ldots, p_{W} ]$ and $\tilde{\mathbf{p}} = [\tilde{p}_1, \ldots, \tilde{p}_{\tilde{W}} ]$ whose elements are the indices corresponding to the non-zero elements in $\mathbf{a}$ and $\tilde{\mathbf{a}}$, respectively, i.e, $\alpha_{p_{w}} \neq 0$ and $\tilde{\alpha}_{\tilde{p}_{\tilde{w}}} \neq 0$. Similarly, define $\mathbf{q} = [q_1, \ldots, q_{Y-W}]$ and $\tilde{\mathbf{q}} = [\tilde{q}_1, \ldots, \tilde{q}_{Y-\tilde{W}}]$ whose elements are the indices corresponding to zero elements in $\mathbf{a}$ and $\tilde{\mathbf{a}}$, respectively, i.e, $\alpha_{q_{w}} = 0$ and $\tilde{\alpha}_{\tilde{q}_{\tilde{w}}} = 0$. Define $\mathbf{\Phi}_{\mathbf{p}}=[\phi_{p_1}, \ldots, \phi_{p_w}]$, $\tilde{\mathbf{\Phi}}_{\tilde{\mathbf{p}}}=[ \tilde{\phi}_{\tilde{p}_{1}}, \ldots, \tilde{\phi}_{\tilde{p}_{\tilde{w}}} ]$, $\mathbf{\Phi}_{\mathbf{q}}=[\phi_{q_1}, \ldots, \phi_{q_w}]$, and $\tilde{\mathbf{\Phi}}_{\tilde{\mathbf{q}}}=[\tilde{\phi}_{\tilde{q}_1}, \ldots, \tilde{\phi}_{\tilde{q}_{\tilde{w}}}]$. Therefore, the PEP 
is written as 
\ifCLASSOPTIONtwocolumn
\begin{align}
\mathrm{Pr} \left( \mathbf{c} \rightarrow \hat{\mathbf{c}} \right) & \leq \mathrm{E} \left[ \exp \left( - \frac{ d^2_{min} (\mathbf{a}^T\mathbf{\Phi}+\tilde{\mathbf{a}}^T\tilde{\mathbf{\Phi}})}{4 N_0} \right) \right] \nonumber \\
& \leq \mathrm{E} \left[ \exp \left( - \mu (\sum^{W}_{w=1} \phi_{p_w} + \sum^{\tilde{W}}_{\tilde{w}=1} \tilde{\phi}_{\tilde{p}_{\tilde{w}}} ) \right) \right]
\label{eq:pep_correlated}
\end{align}
\else
\begin{align}
\mathrm{Pr} \left( \mathbf{c} \rightarrow \hat{\mathbf{c}} \right) \leq \mathrm{E} \left[ \exp \left( - \frac{ d^2_{min} (\mathbf{a}^T\mathbf{\Phi}+\tilde{\mathbf{a}}^T\tilde{\mathbf{\Phi}})}{4 N_0} \right) \right] \leq \mathrm{E} \left[ \exp \left( - \mu (\sum^{W}_{w=1} \phi_{p_w} + \sum^{\tilde{W}}_{\tilde{w}=1} \tilde{\phi}_{\tilde{p}_{\tilde{w}}} ) \right) \right]
\label{eq:pep_correlated}
\end{align}
\fi
with $\mu = \left( d^2_{min} \alpha_{min} \right) / \left( 4 N_0 \right)$, where $\alpha_{min}$ indicates the minimum element in $\mathbf{a}$ and $\tilde{\mathbf{a}}$. To solve (\ref{eq:pep_correlated}), the marginal PDF $f(\mathbf{\Phi}_{\mathbf{p}}, \tilde{\mathbf{\Phi}}_{\tilde{\mathbf{p}}})$ is needed by calculating 
\begin{align}
f(\mathbf{\Phi}_{\mathbf{p}}, \tilde{\mathbf{\Phi}}_{\tilde{\mathbf{p}}}) = \int \cdots \int_{\mathcal{D}_{\mathbf{q}}} \int \cdots \int_{\mathcal{D}_{\tilde{\mathbf{q}}}} f(\mathbf{\Phi}, \tilde{\mathbf{\Phi}}) \, \mathrm{d} \mathbf{\Phi}_{\mathbf{q}} \, \mathrm{d} \tilde{\mathbf{\Phi}}_{\tilde{\mathbf{q}}}.
\label{eq:pdf_marginal}
\end{align}
The joint PDF $f(\mathbf{\Phi}, \tilde{\mathbf{\Phi}})$ is available in \cite{Smith_Wishart, Kuo_Wishart, Chiani_MIMO} as
\begin{align}
f(\mathbf{\Phi}, \tilde{\mathbf{\Phi}}) = \exp \left( - {1 \over 1 - \rho^2} \sum^{Y}_{u=1} (\phi_u + \tilde{\phi}_u ) \right) f_1(\mathbf{\Phi}, \tilde{\mathbf{\Phi}}),
\label{eq:pdf_correlated_wishart}
\end{align}
with the polynomial $f_1(\mathbf{\Phi}, \tilde{\mathbf{\Phi}})$ defined as
\ifCLASSOPTIONtwocolumn
\begin{align}
f_1(\mathbf{\Phi}, \tilde{\mathbf{\Phi}}) & = [ \prod_{u<v}^Y ( \phi_u - \phi_v ) ( \tilde{\phi}_u - \tilde{\phi}_v ) ] \nonumber \\ 
& \quad \times \mathrm{det} [(\phi_u \tilde{\phi}_v)^{(X-Y)/2} I_{X-Y}(2\sqrt{\epsilon \phi_u \tilde{\phi}_v})],
\label{eq:pdf_correlated_wishart_polynomial}
\end{align}
\else
\begin{align}
f_1(\mathbf{\Phi}, \tilde{\mathbf{\Phi}}) = [ \prod_{u<v}^Y ( \phi_u - \phi_v ) ( \tilde{\phi}_u - \tilde{\phi}_v ) ] \times \mathrm{det} [(\phi_u \tilde{\phi}_v)^{(X-Y)/2} I_{X-Y}(2\sqrt{\epsilon \phi_u \tilde{\phi}_v})],
\label{eq:pdf_correlated_wishart_polynomial}
\end{align}
\fi
where $\mathrm{det} [h_{u,v}]$ represents the determinant of the matrix with the $(u,v)$th element given by $h_{u,v}$, $\mathrm{I}_{N}(t)$ is the modified Bessel function of order $N$ which is given by
\begin{align}
I_{N}(t) = \sum_{j=0}^{\infty} {1 \over j!(j+N+1)!}\left({t \over 2}\right)^{2j+N},
\label{eq:modified_bessel}
\end{align}
and $\epsilon \approx \rho^2/(1-\rho^2)^2$. Because the exponent of $\mu = ( d^2_{min} \alpha_{min} ) / ( 4 N_0 )$ is related to the diversity, the constant appearing in the literature is ignored in (\ref{eq:pdf_correlated_wishart_polynomial}) for brevity.

Since the eigenvalues of the Wishart matrices are positive and real, the relations $\exp(-{1 \over 1 - \rho^2}\phi_u) \leq 1$ and $\exp(-{1 \over 1 - \rho^2}\tilde{\phi}_u) \leq 1$ are valid in (\ref{eq:pdf_correlated_wishart}). By applying the relations $\int_{0}^{v} u^t e^{-u} \, \mathrm{d} u \leq {1 \over t+1}v^{t+1}$ and $\int_{0}^{\infty} u^t e^{-u} \, \mathrm{d} u = t!$ to $\mathbf{\Phi}_{\mathbf{q}}$ and $\tilde{\mathbf{\Phi}}_{\tilde{\mathbf{q}}}$, the marginal PDF $f(\mathbf{\Phi}_{\mathbf{p}}, \tilde{\mathbf{\Phi}}_{\tilde{\mathbf{p}}})$ in (\ref{eq:pdf_marginal}) is upper bounded as
\ifCLASSOPTIONtwocolumn
\begin{align}
f(\mathbf{\Phi}_{\mathbf{p}}, \tilde{\mathbf{\Phi}}_{\tilde{\mathbf{p}}}) & \leq \exp \left( -  { \sum^{W}_{w=1} \phi_{p_w} + \sum^{\tilde{W}}_{\tilde{w}=1} \tilde{\phi}_{\tilde{p}_{\tilde{w}}} \over 1 - \rho^2} \right) \nonumber \\ & \times f_2(\mathbf{\Phi}_{\mathbf{p}}, \tilde{\mathbf{\Phi}}_{\tilde{\mathbf{p}}}),
\label{eq:pdf_marginal_polynomial}
\end{align}
\else
\begin{align}
f(\mathbf{\Phi}_{\mathbf{p}}, \tilde{\mathbf{\Phi}}_{\tilde{\mathbf{p}}}) \leq \exp \left( -  { \sum^{W}_{w=1} \phi_{p_w} + \sum^{\tilde{W}}_{\tilde{w}=1} \tilde{\phi}_{\tilde{p}_{\tilde{w}}} \over 1 - \rho^2} \right) \times f_2(\mathbf{\Phi}_{\mathbf{p}}, \tilde{\mathbf{\Phi}}_{\tilde{\mathbf{p}}}),
\label{eq:pdf_marginal_polynomial}
\end{align}
\fi
where $f_2(\mathbf{\Phi}_{\mathbf{p}}, \tilde{\mathbf{\Phi}}_{\tilde{\mathbf{p}}})$ is a polynomial corresponding to (\ref{eq:pdf_correlated_wishart}). Then (\ref{eq:pep_correlated}) is rewritten as 
\ifCLASSOPTIONtwocolumn
\begin{align}
\mathrm{Pr} \left( \mathbf{c} \rightarrow \hat{\mathbf{c}} \right) & \leq \int_{0}^{\infty} \int_{0}^{\phi_{p_{1}}} \cdots \int_{0}^{\phi_{p_{W-1}}} \int_{0}^{\infty} \int_{0}^{\phi_{\tilde{p}_{1}}} \cdots \int_{0}^{\phi_{\tilde{p}_{\tilde{W}-1}}} \nonumber \\ 
& \,\, \exp \left( - ( \mu + {1 \over 1 - \rho^2} ) (\sum^{W}_{w=1} \phi_{p_w} + \sum^{\tilde{W}}_{\tilde{w}=1} \tilde{\phi}_{\tilde{p}_{\tilde{w}}} ) \right) \nonumber \\ 
& \,\, \times f_2(\mathbf{\Phi}_{\mathbf{p}}, \tilde{\mathbf{\Phi}}_{\tilde{\mathbf{p}}}) \, \mathrm{d} \mathbf{\Phi}_{\mathbf{p}} \, \mathrm{d} \tilde{\mathbf{\Phi}}_{\tilde{\mathbf{p}}}.
\label{eq:pep_correlated_polynomial}
\end{align}
\else
\begin{align}
\mathrm{Pr} \left( \mathbf{c} \rightarrow \hat{\mathbf{c}} \right) & \leq \int_{0}^{\infty} \int_{0}^{\phi_{p_{1}}} \cdots \int_{0}^{\phi_{p_{W-1}}} \int_{0}^{\infty} \int_{0}^{\phi_{\tilde{p}_{1}}} \cdots \int_{0}^{\phi_{\tilde{p}_{\tilde{W}-1}}} \nonumber \\ 
& \quad \exp \left( - ( \mu + {1 \over 1 - \rho^2} ) (\sum^{W}_{w=1} \phi_{p_w} + \sum^{\tilde{W}}_{\tilde{w}=1} \tilde{\phi}_{\tilde{p}_{\tilde{w}}} ) \right) \times f_2(\mathbf{\Phi}_{\mathbf{p}}, \tilde{\mathbf{\Phi}}_{\tilde{\mathbf{p}}}) \, \mathrm{d} \mathbf{\Phi}_{\mathbf{p}} \, \mathrm{d} \tilde{\mathbf{\Phi}}_{\tilde{\mathbf{p}}}.
\label{eq:pep_correlated_polynomial}
\end{align}
\fi
Note that since $f_2(\mathbf{\Phi}_{\mathbf{p}}, \tilde{\mathbf{\Phi}}_{\tilde{\mathbf{p}}})$ is a polynomial, its multivariate terms can be integrated separately, and the term with the worst performance dominates the overall performance. To solve (\ref{eq:pep_correlated_polynomial}), Theorem $2$ in \cite{Park_UP_MPDF} can be applied to integrate $\mathbf{\Phi}_{\mathbf{p}}$ and $\tilde{\mathbf{\Phi}}_{\tilde{\mathbf{p}}}$ independently for each multivariate term. 
Based on this theorem 
the multivariate term with the smallest degree in $f_2(\mathbf{\Phi}_{\mathbf{p}}, \tilde{\mathbf{\Phi}}_{\tilde{\mathbf{p}}})$ results in the smallest degree of $(\mu+{1 \over 1- \rho^2})^{-1}$, which dominates the overall performance. The smallest degree of $f_2(\mathbf{\Phi}_{\mathbf{p}}, \tilde{\mathbf{\Phi}}_{\tilde{\mathbf{p}}})$ is $(X-p_1+1)(Y-p_1+1)+(X-\tilde{p}_1+1)(Y-\tilde{p}_1+1)-W-\tilde{W}$, and the proof is provided in Appendix A. Therefore, (\ref{eq:pep_correlated_polynomial}) is upper bounded by 
\begin{align}
\mathrm{Pr} \left( \mathbf{c} \rightarrow \hat{\mathbf{c}} \right) & \leq \zeta\left({ d^2_{min} \alpha_{min} \over 4N_t}\gamma+{1 \over 1 - \rho^2}\right) ^{-D},
\label{eq:pep_diversity_correlated}
\end{align}
with $D=(X-p_1+1)(Y-p_1+1)+(X-\tilde{p}_1+1)(Y-\tilde{p}_1+1)$, where $\zeta$ is a constant.

The negative effect of subcarrier correlation is proved by (\ref{eq:pep_diversity_correlated}). When $\gamma \to \infty$, the diversity order is the same as the uncorrelated case in (\ref{eq:pep_diversity}) and (\ref{eq:diversity}). However, on the practical SNR range, the performance is degraded due to the term $1 \over 1 - \rho^2$, which is independent of $\gamma$. Specifically, when the subcarrier correlation $\rho$ is small, $1 \over 1 - \rho^2$ is also relatively small, and its effect on the performance is not significant when the SNR is relatively large, and the uncorrelated case $\rho = 0$ offers the performance upper bound. On the other hand, when $\rho$ is large, $1 \over 1 - \rho^2$ is also relatively large compared to $\gamma$, then significant performance degradation could be caused, depending on the SNR. When $\rho = 1$, which means all the distinct bits of the error path are transmitted through one subcarrier, no multipath diversity is achieved, and the diversity equals BICMB in the case of flat fading MIMO channels introduced in \cite{Park_DA_BICMB, Park_DA_BICMB_J}, which provides the performance lower bound. 

Note that the analysis in this section is not limited to equal power channel taps, and can also be applied to unequal power channel taps, non-constant sampling time, and other assumptions, which cause different subcarrier correlation.

%% file: Grouping.tex
\section{Subcarrier Grouping} \label{sec:grouping}

The idea of subcarrier grouping technique is to transmit multiple streams of information through multiple group of subcarriers of OFDM. It was suggested for multi-user interference elimination \cite{Wang_WMC}, Peak-to-Average Ratio (PAR) reduction \cite{Goeckel_OFDM}, and complexity reduction \cite{Liu_LCP_OFDM}. In this paper, the subcarrier correlation technique is applied to overcome the performance loss caused by subcarrier correlation. 

Note that $\rho = 0$ 
when $(m-m')L/M \in \mathbb{Z}$, where $\mathbb{Z}$ denotes the set of integer numbers. This means that although correlation exists among subcarriers for $L<M$, some subcarriers could be uncorrelated if $M/L \in \mathbb{Z}^+$, where $\mathbb{Z}^+$ denotes the set of positive integer numbers. In this case, there are $G=M/L$ groups of $L$ uncorrelated subcarriers. As a result, the subcarrier grouping technique can be applied to transmit multiple streams of bit codewords through these $G$ different groups of uncorrelated subcarriers, instead of transmitting one stream of the bit codeword through all the correlated subcarriers. As a result, the negative effect of subcarrier correlation is completely avoided, and the maximum achievable diversity is thereby achieved. Note that the best choice of the number of subcarriers in one group is $L$, since smaller choice results in less diversity while larger choice causes subcarrier correlation which degrades performance.
\\\textbf{Example:} Consider the case of $L=2$ and $M=64$. Then, the $g$th and the $(g+32)$th subcarriers are uncorrelated for $g=1,\ldots,32$. The subcarrier grouping technique can transmit $32$ streams of bit codewords simultaneously through the $32$ groups of two uncorrelated subcarriers without performance degradation. 

\ifCLASSOPTIONonecolumn
\begin{figure}[!t]
\centering \includegraphics[width = 1.0\linewidth]{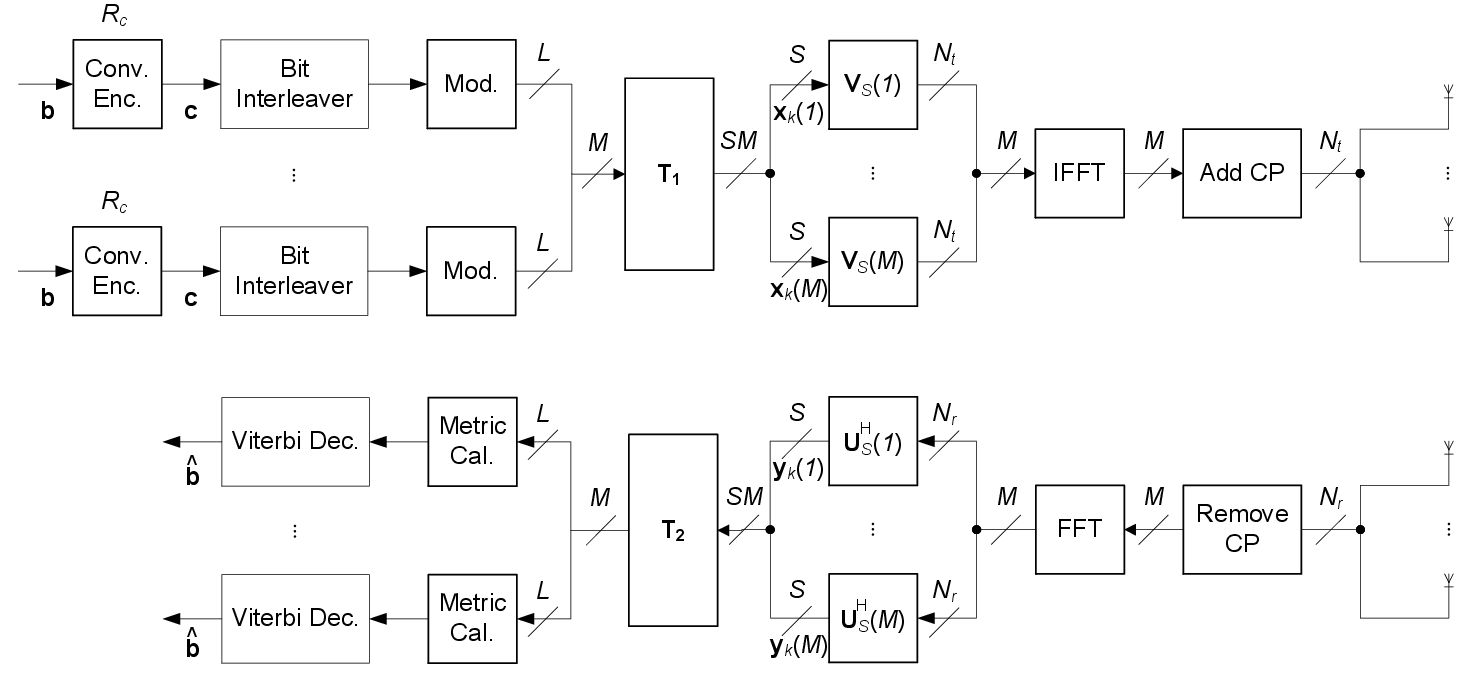}
\caption{Structure of BICMB-OFDM with subcarrier grouping.} \label{fig:bicmb_ofdm_sg}
\end{figure}
\else
\begin{figure}[!t]
\centering \includegraphics[width = 1.0\linewidth]{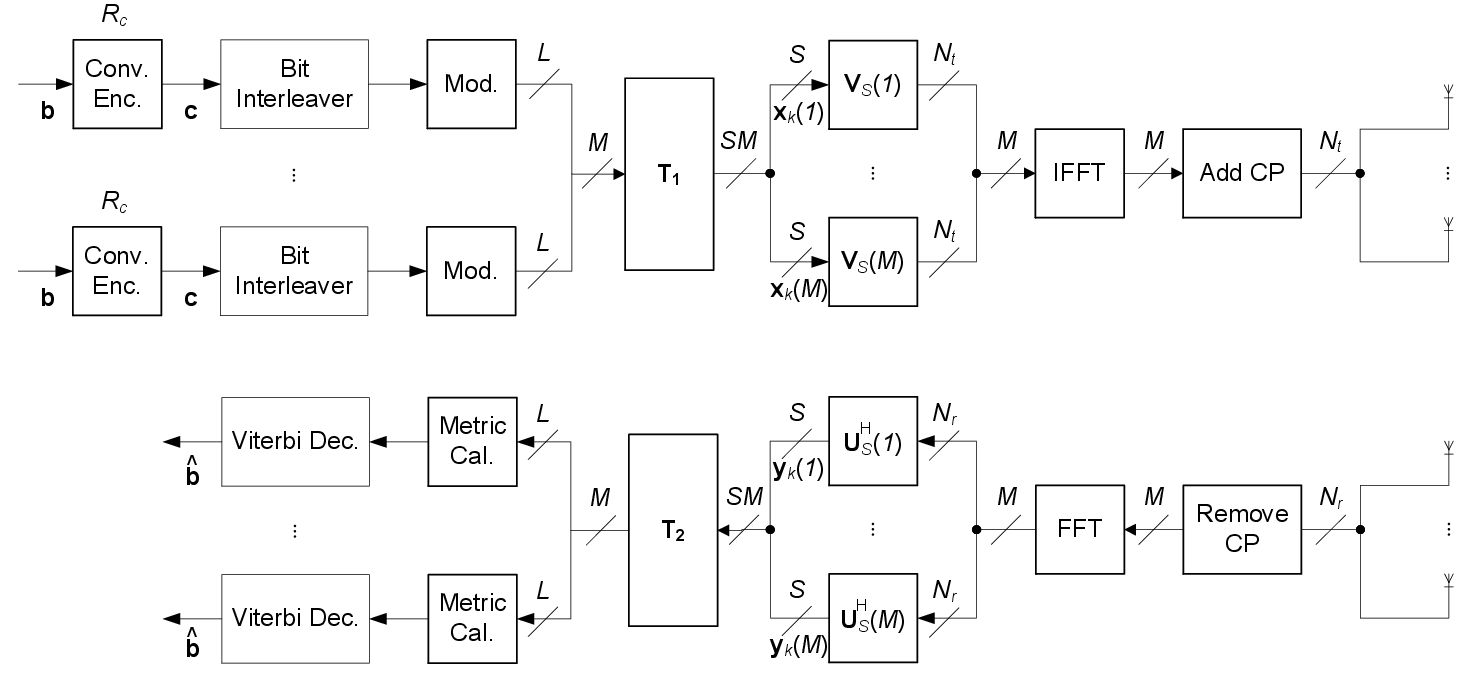}
\caption{Structure of BICMB-OFDM with subcarrier grouping.} \label{fig:bicmb_ofdm_sg}
\end{figure}
\fi

Fig. \ref{fig:bicmb_ofdm_sg} presents the structure of BICMB-OFDM with subcarrier grouping. In Fig. \ref{fig:bicmb_ofdm_sg}, $\mathbf{T}_1$ is a permutation matrix at the transmitter distributing the modulated symbols from different streams to their corresponding subcarriers, while $\mathbf{T}_2=\mathbf{T}_1^{-1}$ is a permutation matrix at the receiver distributing the received symbols of different subcarriers to their corresponding streams for decoding. Compared to BICMB-OFDM without subcarrier grouping, BICMB-OFDM with subcarrier grouping achieves better performance with the same transmission rate and decoding complexity. Note that the structure of BICMB-OFDM with subcarrier grouping in Fig. \ref{fig:bicmb_ofdm_sg} can also be considered as Orthogonal Frequency-Division Multiple Access (OFDMA) \cite{Ghosh_LTE} version of BICMB-OFDM. OFDMA is a multi-user version of the OFDM and it has been used in the mobility mode of WiMAX \cite{IEEE_802_16} as well as the downlink of LTE \cite{3GPP_TS_36.201}. The multiple access in OFDMA is achieved by assigning subsets of subcarriers to individual users, which is similar to the subcarrier grouping technique. As a result, with subcarrier grouping, BICMB-OFDM can provide multi-user compatibility. 

\ifCLASSOPTIONonecolumn
\begin{figure}[!t]
\centering \includegraphics[width = 1.0\linewidth]{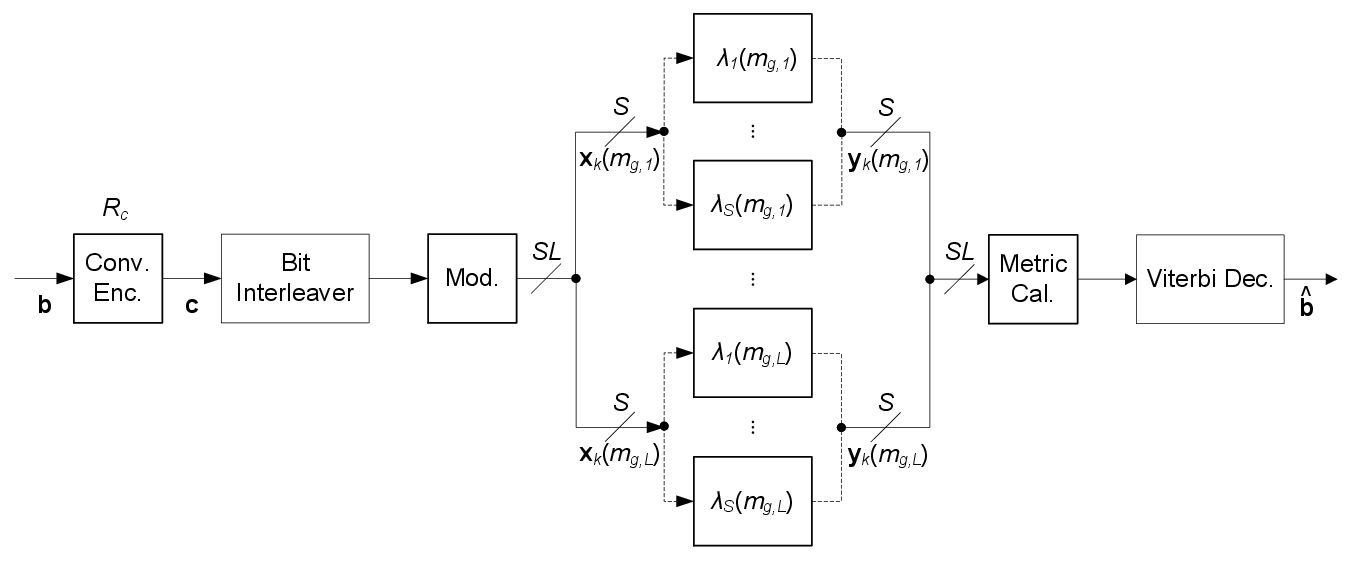}
\caption{Structure of BICMB-OFDM with subcarrier grouping in the frequency domain for one bit stream transmission of the $g$th subcarrier group.} \label{fig:bicmb_ofdm_sgf}
\end{figure}
\else
\begin{figure}[!t]
\centering \includegraphics[width = 1.0\linewidth]{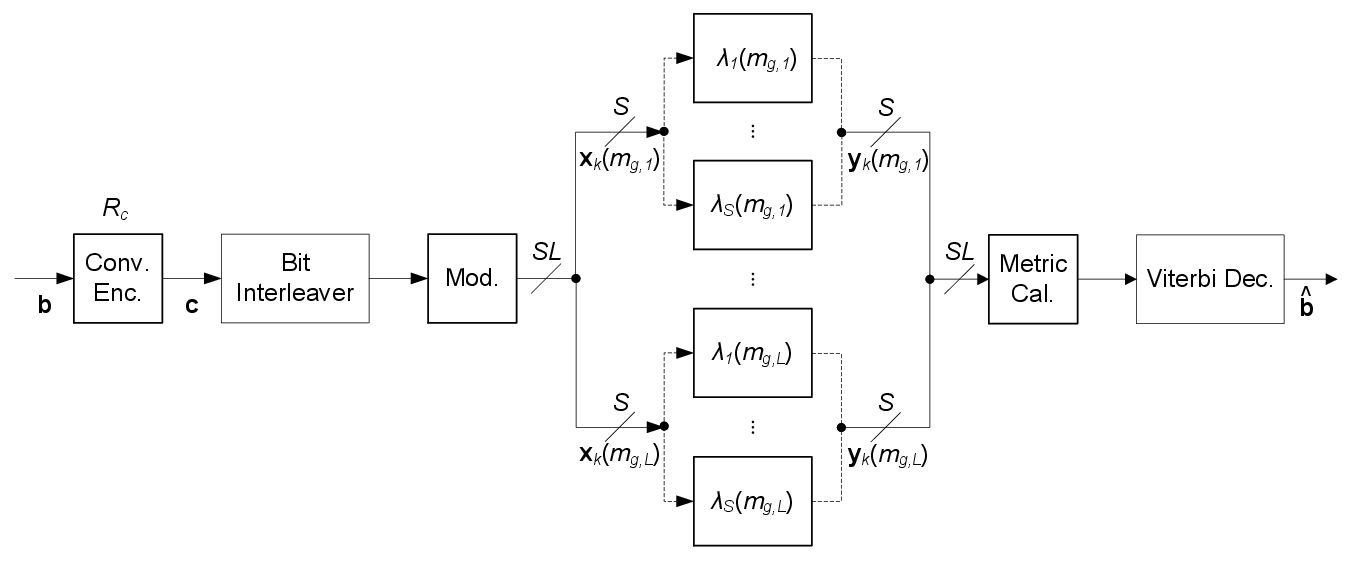}
\caption{Structure of BICMB-OFDM with subcarrier grouping in the frequency domain for one bit stream transmission of the $g$th subcarrier group.} \label{fig:bicmb_ofdm_sgf}
\end{figure}
\fi

Fig. \ref{fig:bicmb_ofdm_sgf} presents the structure of BICMB-OFDM with subcarrier grouping in the frequency domain for one bit stream transmission of the $g$th subcarrier group with $g \in\{1, \ldots, G \}$, and $m_{g,l}=(l-1)G+g$ for $l=1,\ldots,L$ in the figure denotes the corresponding subcarrier index for the $l$th subcarrier of the $g$th group. Note that Fig. \ref{fig:bicmb_ofdm_sgf} can also present the structure of BICMB-OFDM in the frequency domain when $L=M$. Therefore, the diversity analysis for $L=M$ in Section \ref{sec:diversity} can also be applied to BICMB-OFDM with subcarrier grouping. As a result, the full diversity condition $R_cSL \leq 1$ holds for BICMB-OFDM with subcarrier grouping as well. In this paper, the number of employed subchannels by SVD for each subcarrier is assumed to be the same, which is $S$. However, they could be different in practice. In that case, the full diversity condition is $R_c\sum_{l=1}^{L}S_{g,l} \leq 1$ where $S_{g,l}$ denotes the number of employed subchannels by SVD for the $l$th subcarrier of the $g$th group.

Note that when the channel taps have different powers, there are no uncorrelated subcarriers in general \cite{Li_DA_BICMB_OFDM, Li_DA_BICMB_OFDM_arXiv}. However, some of them could have weak correlation. Therefore, the subcarrier grouping technique can still be applied to combat the performance degradation, although it now can no longer fully recover the performance because of subcarrier correlation. 

%% file: Results.tex
\section{Simulation Results} \label{sec:results}

To verify the diversity analysis, $2 \times 2$, $M=64$ BICMB-OFDM with $L=2$ and $L=4$ using $4$-QAM are considered for simulations. The number of employed subchannels for each subcarrier is assumed to be the same. The generator polynomials in octal for the convolutional codes with $R_c=1/4$ and $R_c= 1/2$ are $(5,7,7,7)$, and $(5,7)$ respectively, and the codes with $R_c = 2/3$ and $R_c = 4/5$ are punctured from the $R_c=1/2$ code \cite{Haccoun_PCC}. The length of CP is $L_{cp}=16$. Each OFDM symbol has $4\mu \mathrm{s}$ duration, of which $0.8\mu \mathrm{s}$ is CP. Equal and exponential power channel taps are considered. For the exponential channel model \cite{Cho_MIMO_OFDM}, the ratios of non-negligible path power to the first path power are $-7\mathrm{dB}$, the mean excess delays are $30n\mathrm{s}$ for $L=2$ and $65n\mathrm{s}$ for $L=4$, respectively. The bit interleaver employs simple rotation. Note that simulations of $2 \times 2$, $L=2$ and $L=4$ BICMB-OFDM are shown in this section because the diversity values could be investigated explicitly through figures. 

\subsection{Diversity of BICMB-OFDM with Subcarrier Grouping} \label{subsec:results_sg}

\ifCLASSOPTIONonecolumn
\begin{figure}[!t]
\centering \includegraphics[width = 1.0\linewidth]{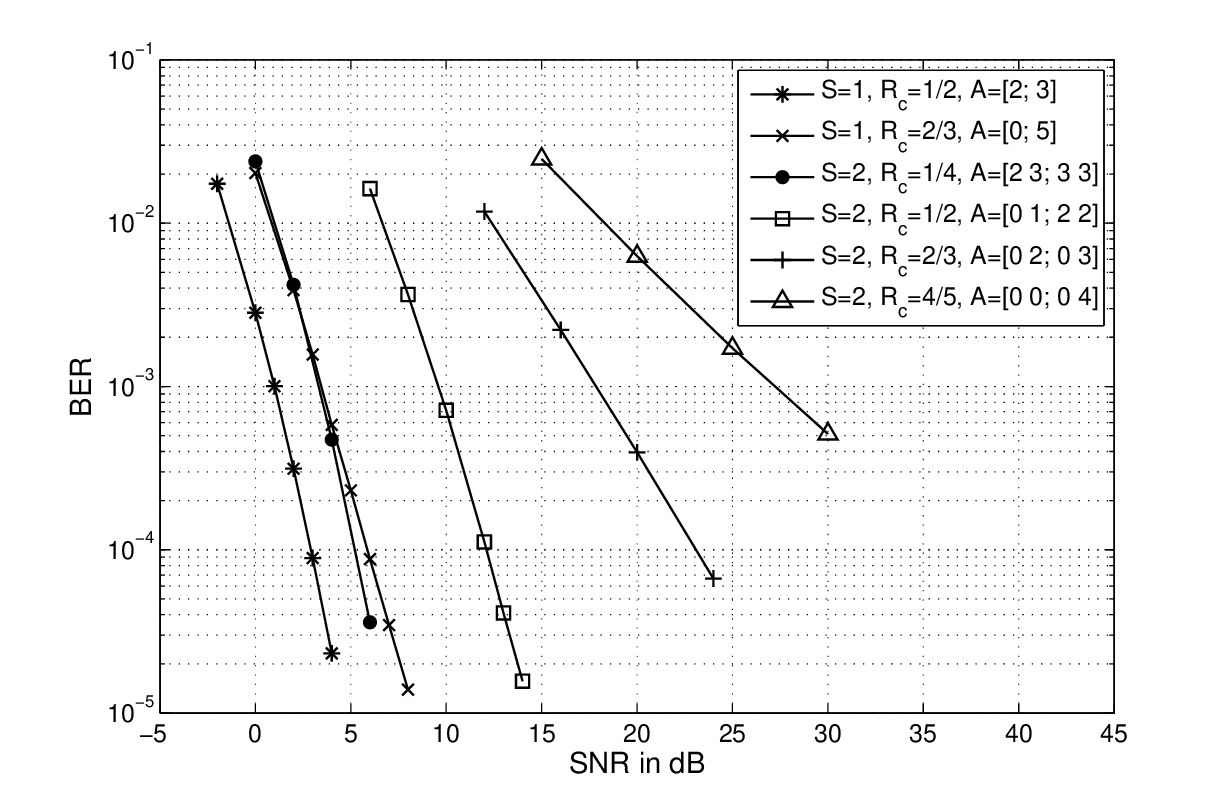}
\caption{BER vs. SNR for $2 \times 2$, $L=2$, $M=64$ BICMB-OFDM with subcarrier grouping over equal power channel taps.}
\label{fig:uncorrelated_2tap}
\end{figure}
\else
\begin{figure}[!t]
\centering \includegraphics[width = 1.0\linewidth]{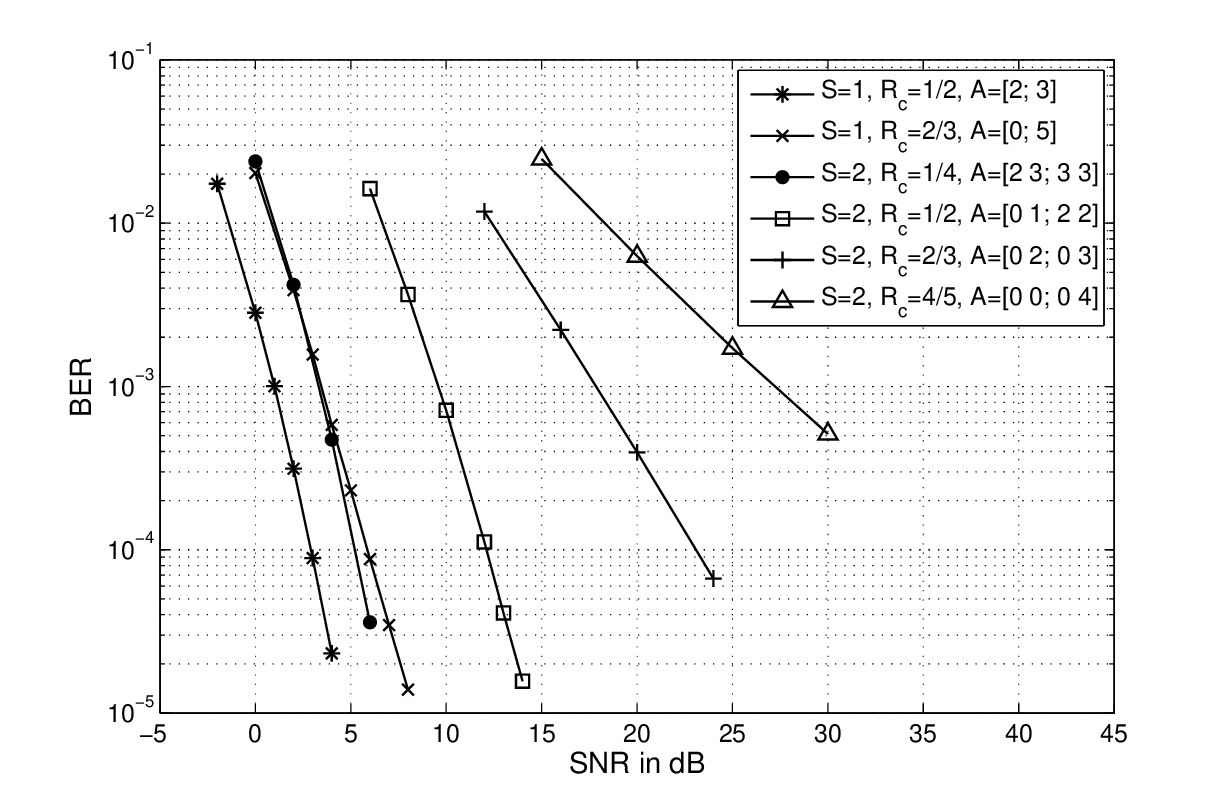}
\caption{BER vs. SNR for $2 \times 2$, $L=2$, $M=64$ BICMB-OFDM with subcarrier grouping over equal power channel taps.}
\label{fig:uncorrelated_2tap}
\end{figure}
\fi

Fig. \ref{fig:uncorrelated_2tap} shows the Bit Error Rate (BER) performance of $2 \times 2$, $L=2$, $M=64$ BICMB-OFDM employing subcarrier grouping over equal power channel taps with different $S$ and $R_c$. The $\mathbf{A}$ matrices that dominate the performance derived by the method introduced in Section \ref{subsec:alpha} are provided in the figure. The diversity results of all curves equal the maximum achievable diversity orders derived from Section \ref{sec:diversity}, which are directly decided by the $\mathbf{A}$ matrices. Specifically, in the cases of $S=1$, $R_c=1/4$ and $R_c=2/3$ codes, whose dominant $\mathbf{A}$ matrices are $\mathbf{A} = [2 ; \, 3]$ and $\mathbf{A} = [0 ;\, 5]$ respectively, achieve diversity values of $8$ and $4$ respectively. As for $S=2$, the codes with $R_c=1/4$, $R_c=1/2$, $R_c=2/3$, and $R_c=4/5$, whose dominant $\mathbf{A}$ matrices are $\mathbf{A} = [2 \, 3; 3 \, 3]$, $\mathbf{A} = [0 \, 1; 2 \, 2]$, $\mathbf{A} = [0 \, 2; 0 \, 3]$, and $\mathbf{A} = [0 \, 0; 0 \, 4]$ respectively, offer diversity of $8$, $5$, $2$, and $1$ respectively. Note that full diversity of $8$ is achieved with the condition $R_cSL \leq 1$.

\ifCLASSOPTIONonecolumn
\begin{figure}[!t]
\centering \includegraphics[width = 1.0\linewidth]{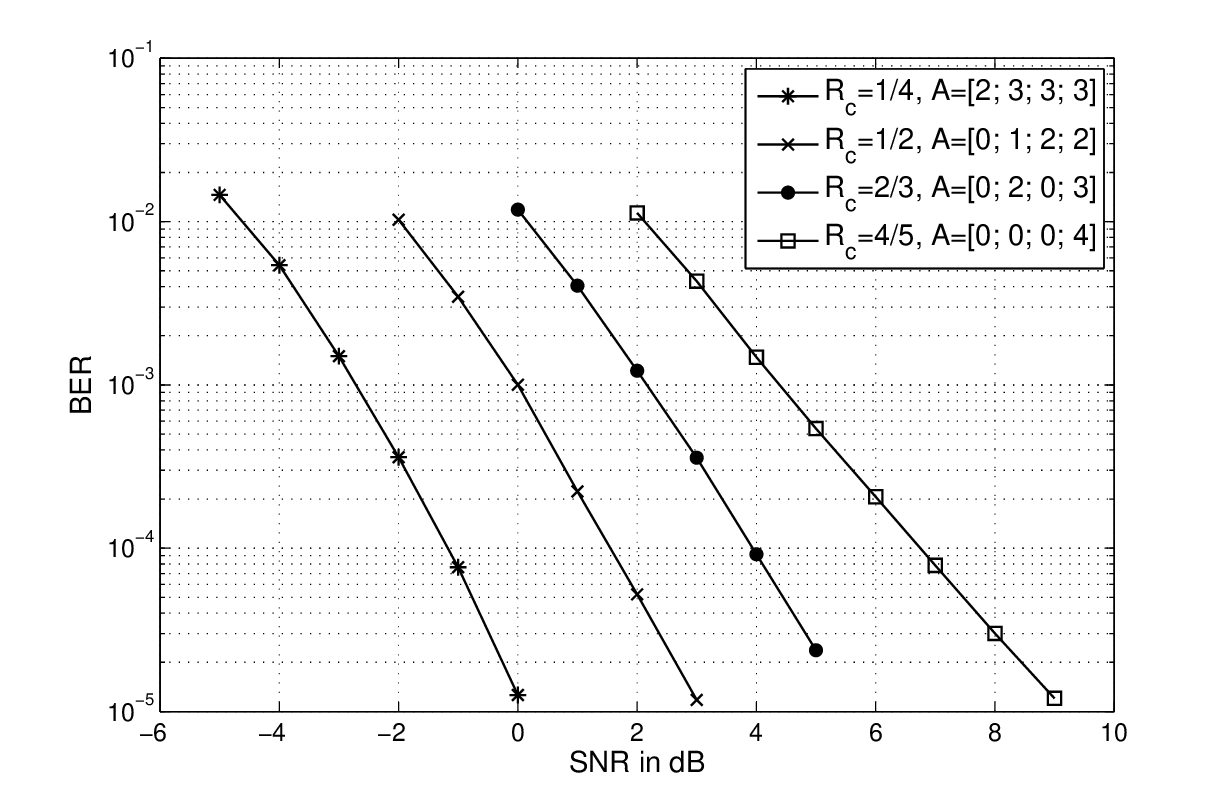}
\caption{BER vs. SNR for $2 \times 2$, $L=4$, $M=64$, $S=1$ BICMB-OFDM with subcarrier grouping over equal power channel taps.}
\label{fig:uncorrelated_4tap}
\end{figure}
\else
\begin{figure}[!t]
\centering \includegraphics[width = 1.0\linewidth]{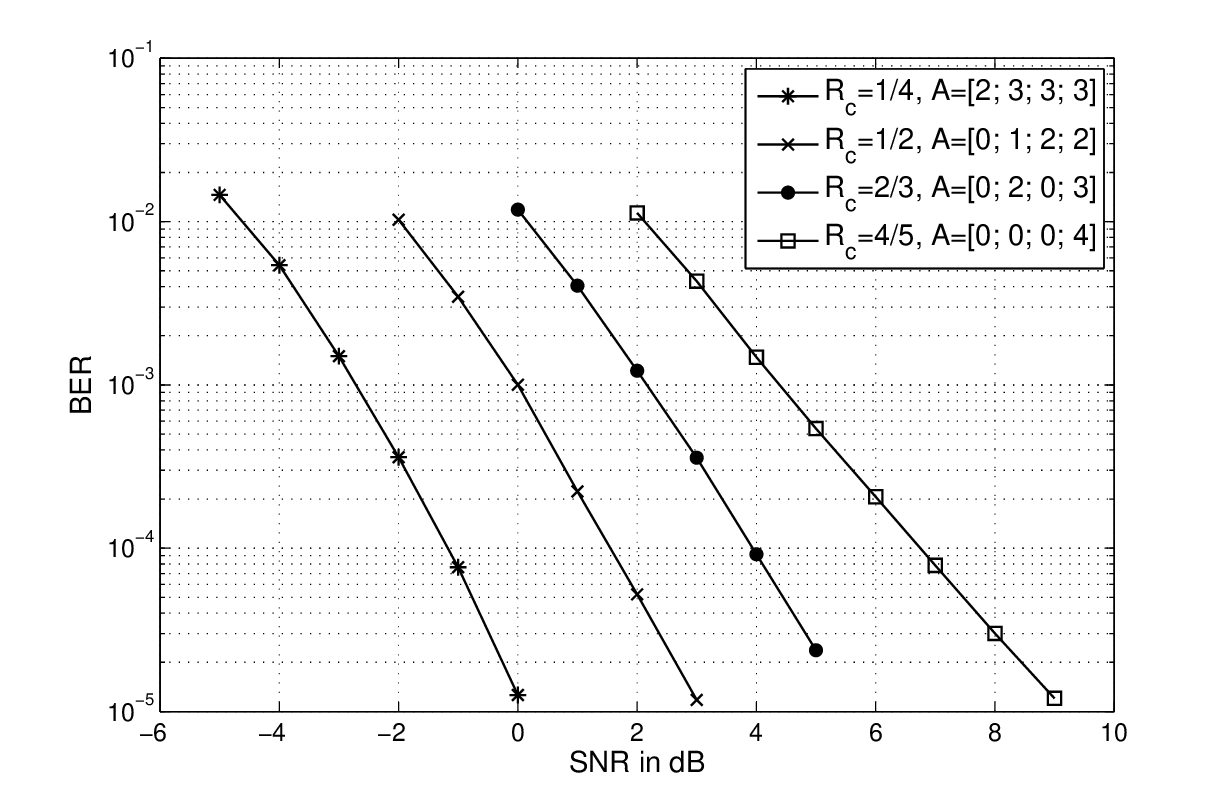}
\caption{BER vs. SNR for $2 \times 2$, $L=4$, $M=64$, $S=1$ BICMB-OFDM with subcarrier grouping over equal power channel taps.}
\label{fig:uncorrelated_4tap}
\end{figure}
\fi

Similarly, Fig. \ref{fig:uncorrelated_4tap} shows the BER performance of $2 \times 2$, $L=4$, $M=64$, $S=1$ BICMB-OFDM employing subcarrier grouping over equal power channel taps with different $R_c$. The $\mathbf{A}$ matrices that dominate the performance derived by the method introduced in Section \ref{subsec:alpha} are provided in the figure. The diversity results of all curves equal the maximum achievable diversity orders derived from Section \ref{sec:diversity}, which are directly decided by the $\mathbf{A}$ matrices. Specifically, the codes with $R_c=1/4$, $R_c=1/2$, $R_c=2/3$, and $R_c=4/5$, whose dominant $\mathbf{A}$ matrices are $\mathbf{A} = [2;\, 3;\, 3;\, 3]$, $\mathbf{A} = [0;\, 1;\, 2;\, 2]$, $\mathbf{A} = [0;\, 2;\, 0;\, 3]$, and $\mathbf{A} = [0;\, 0;\, 0;\, 4]$ respectively, offer diversity of $16$, $12$, $8$, and $4$ respectively. Note that full diversity of $16$ is achieved with the condition $R_cSL \leq 1$.

Fig. \ref{fig:uncorrelated_2tap} and Fig. \ref{fig:uncorrelated_4tap} verify the relation between the diversity and the $\alpha$-spectra as well as the full diversity condition $R_cSL \leq 1$ derived in Section \ref{sec:diversity} for BICMB-OFDM with subcarrier grouping. The full diversity condition implies that if the number of streams $S$ transmitted at each subcarrier increases, the code rate $R_c$ may have to decrease in order to keep full diversity. As a result, increasing the number of parallel streams may not fully improve the total transmission rate, which is a similar issue to the full diversity condition $R_cS \leq 1$ of BICMB for flat fading MIMO channels introduced in \cite{Park_DA_BICMB, Park_DA_BICMB_J}. In fact, for flat fading MIMO channels, other than channel coding, the constellation precoding technique has been incorporated with both uncoded and coded SVD beamforming to achieve full diversity and full multiplexing simultaneously, with the trade-off of a higher decoding complexity \cite{Park_CPB, Park_BICMB_CP, Park_MB_CP, Park_CPMB}. Specifically, in the uncoded case, full diversity requires that all streams are precoded. On the other hand, for the coded case, which is BICMB, even without the condition $R_cS \leq 1$, other than full precoding, partial precoding with lower decoding complexity than full precoding could also achieve both full diversity and full multiplexing with the properly designed combination of the convolutional code, the bit interleaver, and the constellation precoder. Moreover, Perfect Space-Time Block Codes (PSTBCs) \cite{Oggier_PSTBC}, which have the properties of full rate, full diversity, uniform average transmitted energy per antenna, good shaping of the constellation, and nonvanishing constant minimum determinant for increasing spectral efficiency which offers high coding gain, have been considered as an alternative scheme to replace the constellation precoding technique for both uncoded and coded SVD beamforming with constellation precoding in order to reduce the decoding complexity in dimensions $2$ and $4$ while achieving almost the same performance \cite{Li_GCMB, Li_BICMB_PC, Li_MB_PC}. Since these techniques have successfully solved the restricted full diversity condition issue of $R_cS \leq 1$ for BICMB in the case of flat fading MIMO channels, it may be possible to incorporate these techniques into BICMB-OFDM so that its full diversity condition is not restricted to $R_cSL \leq 1$ for frequency selective MIMO channels. However, the design criteria and diversity analysis cannot be generalized in a straightforward manner because of the increased system complexity, and they are discussed in another work by the authors \cite{Li_FDPD_BICMB_OFDM_J, Li_FDPD_BICMB_OFDM}.

\subsection{Negative Effect of Subcarrier Correlation} \label{subsec:results_correlation}

\ifCLASSOPTIONonecolumn
\begin{figure}[!t]
\centering \includegraphics[width = 1.0\linewidth]{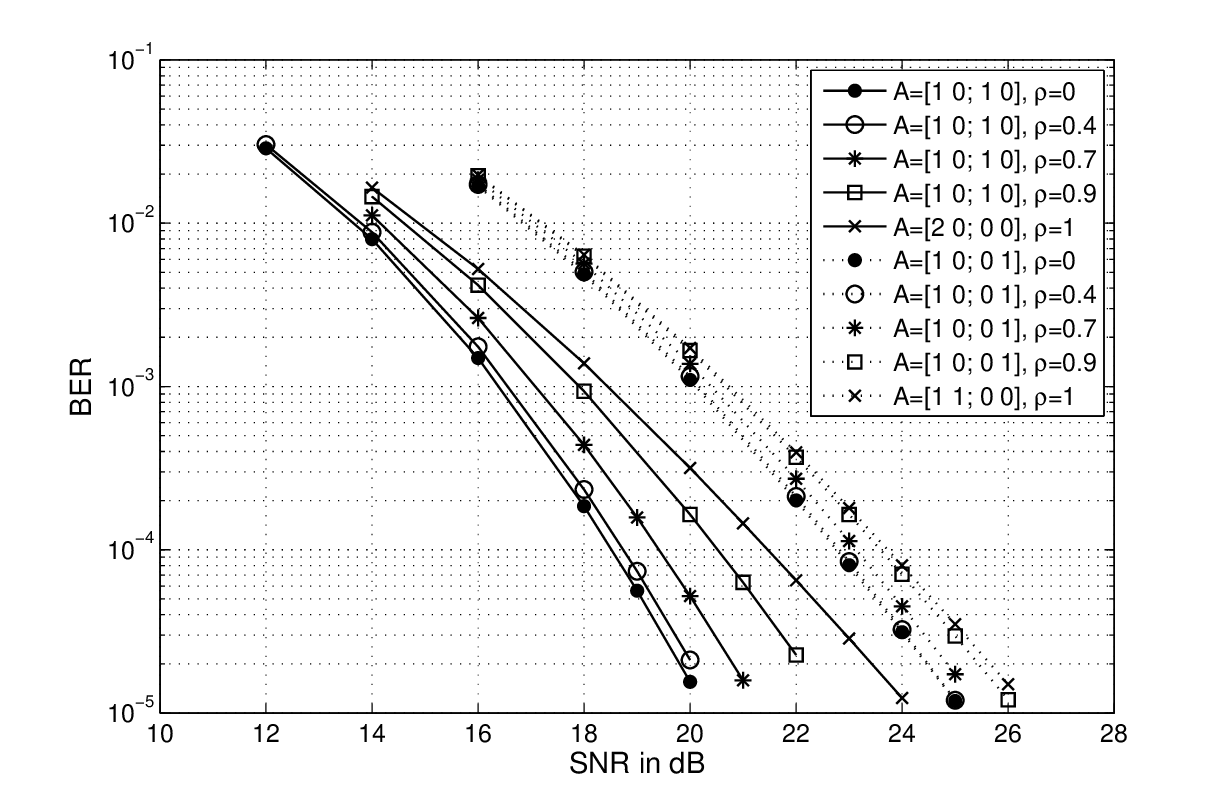}
\caption{BER vs. SNR for examined PEPs of two subcarriers with different correlation coefficient for $2 \times 2$, $L=2$, $M=64$, $S=2$ BICMB-OFDM over equal power channel taps.}
\label{fig:expectation}
\end{figure}
\else
\begin{figure}[!t]
\centering \includegraphics[width = 1.0\linewidth]{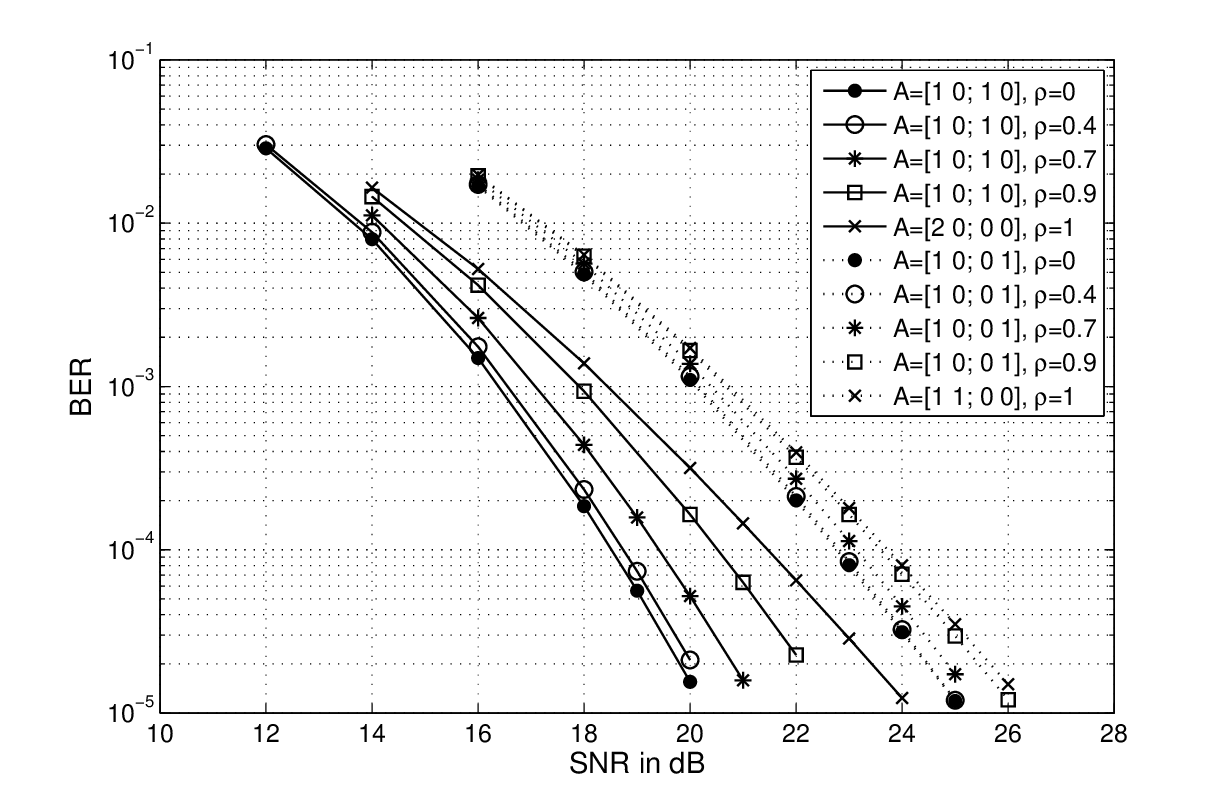}
\caption{BER vs. SNR for examined PEPs of two subcarriers with different correlation coefficient for $2 \times 2$, $L=2$, $M=64$, $S=2$ BICMB-OFDM over equal power channel taps.}
\label{fig:expectation}
\end{figure}
\fi

Fig. \ref{fig:expectation} shows the BER performance of examined PEPs 
with $S=2$, where the simplest case of an error event with $d_H=2$ is examined for two subcarriers with different correlation coefficient $\rho$ in absolute value, which are derived from the $2 \times 2$, $L=2$, $M=64$ BICMB-OFDM over equal power channel taps. The figure shows that when $\rho = 0$, which implies that the two subcarriers are uncorrelated, $\mathbf{A} = [1 \, 0; 1 \, 0]$ and $\mathbf{A} = [1 \, 0; 0 \, 1]$ offer diversity of $8$ and $5$ respectively. On the other hand, when $\rho \neq 0$, performance degradation is caused by subcarrier correlation, and stronger correlation results in worse performance loss. When $\rho = 1$, which means that the $d_H=2$ distinct bits are transmitted through only one subcarrier and no multipath diversity is achieved, both $\mathbf{A} = [2 \, 0; 0 \, 0]$ and $\mathbf{A} = [1 \, 1; 0 \, 0]$ provide diversity of $4$. The results are consistent with the analysis provided in Section \ref{sec:correlation}, and they show the negative effect of subcarrier correlation on performance.



\ifCLASSOPTIONonecolumn
\begin{figure}[!t]
\centering \includegraphics[width = 1.0\linewidth]{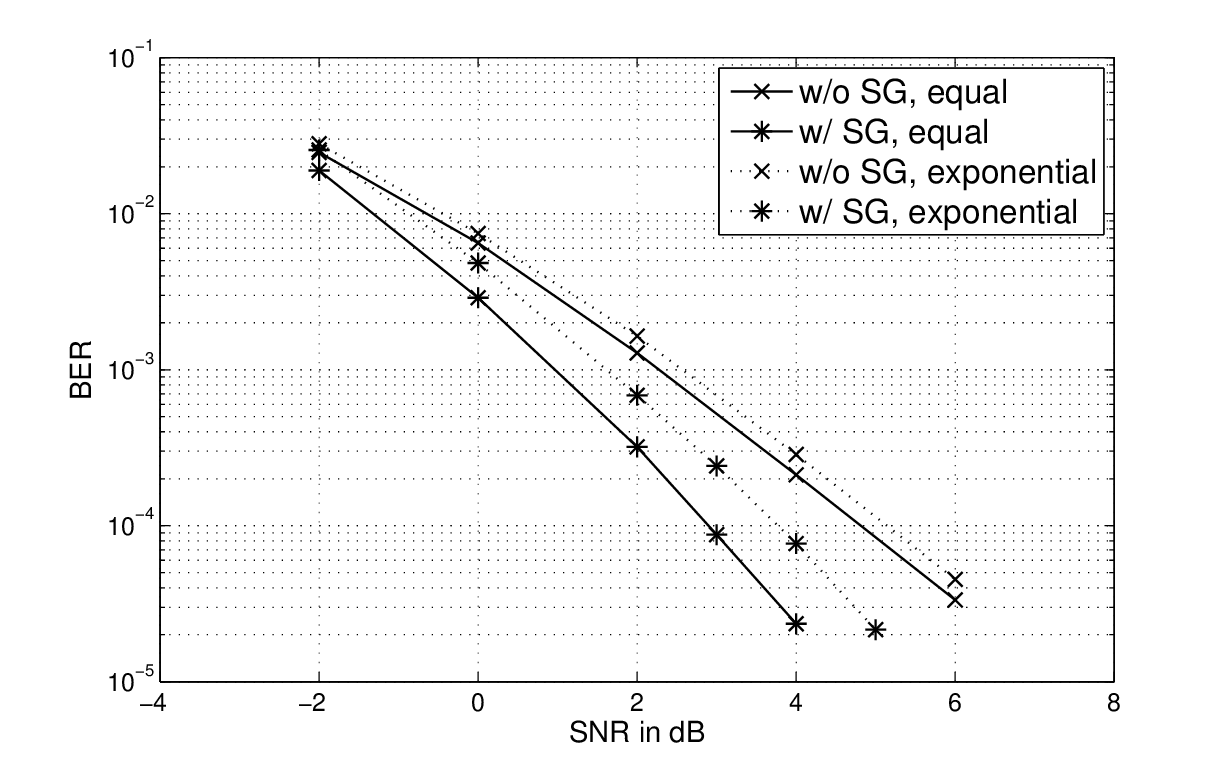}
\caption{BER vs. SNR for $2 \times 2$, $L=2$, $M=64$, $S=1$, $R_c=1/2$ BICMB-OFDM with and without subcarrier grouping over equal and exponential power channel taps.}
\label{fig:exp_2tap}
\end{figure}
\else
\begin{figure}[!t]
\centering \includegraphics[width = 1.0\linewidth]{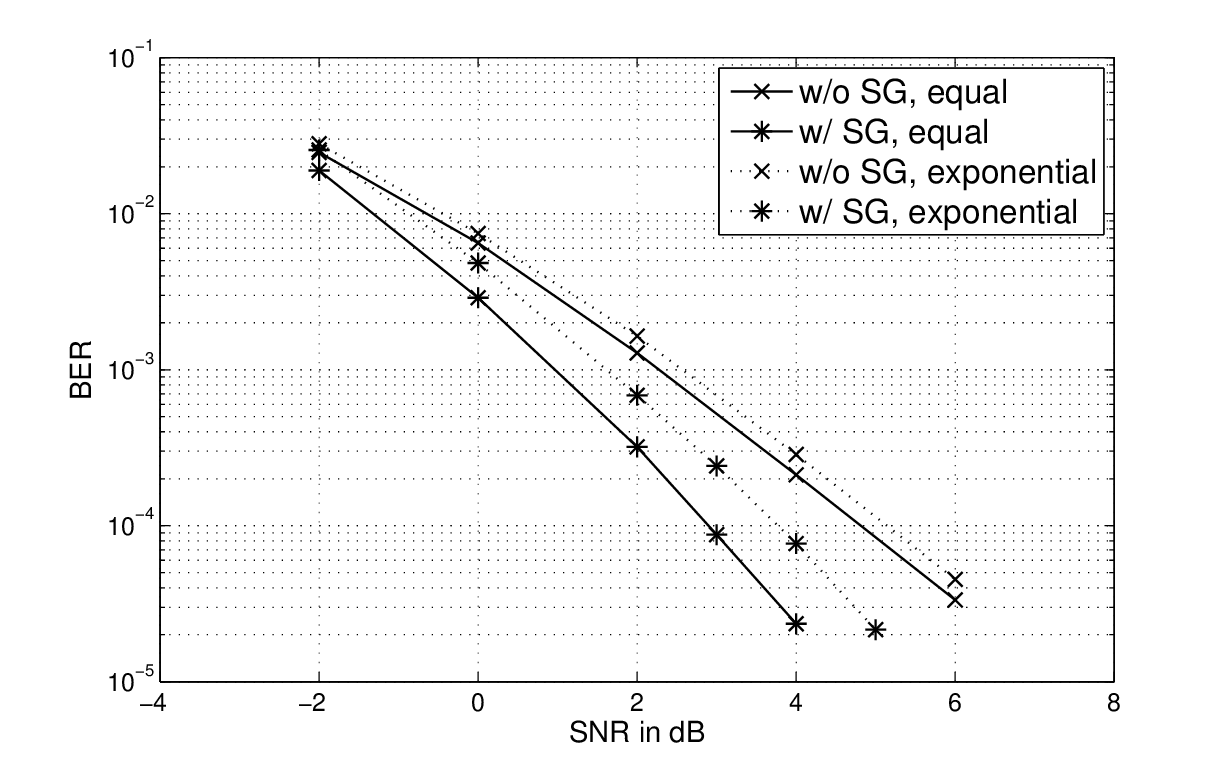}
\caption{BER vs. SNR for $2 \times 2$, $L=2$, $M=64$, $S=1$, $R_c=1/2$ BICMB-OFDM with and without subcarrier grouping over equal and exponential power channel taps.}
\label{fig:exp_2tap}
\end{figure}
\fi

Fig. \ref{fig:exp_2tap} shows the BER performance of $2 \times 2$, $L=2$, $M=64$, $S=1$, $R_c=1/2$ BICMB-OFDM with and without subcarrier grouping over equal and exponential power channel taps. In the figure, w/ and w/o denote with and without respectively, while SG denotes subcarrier grouping. The results show that the subcarrier grouping technique can combat the performance loss caused by subcarrier correlation for both equal and exponential power channel taps. As discussed in Section \ref{sec:grouping}, the maximum achievable diversity of $8$ is provided by employing subcarrier grouping for equal power channel taps, since there is no subcarrier correlation. As for the case of exponential power channel taps, subcarrier grouping cannot fully recover the performance loss because subcarrier correlation still exists. 

\ifCLASSOPTIONonecolumn
\begin{figure}[!t]
\centering \includegraphics[width = 1.0\linewidth]{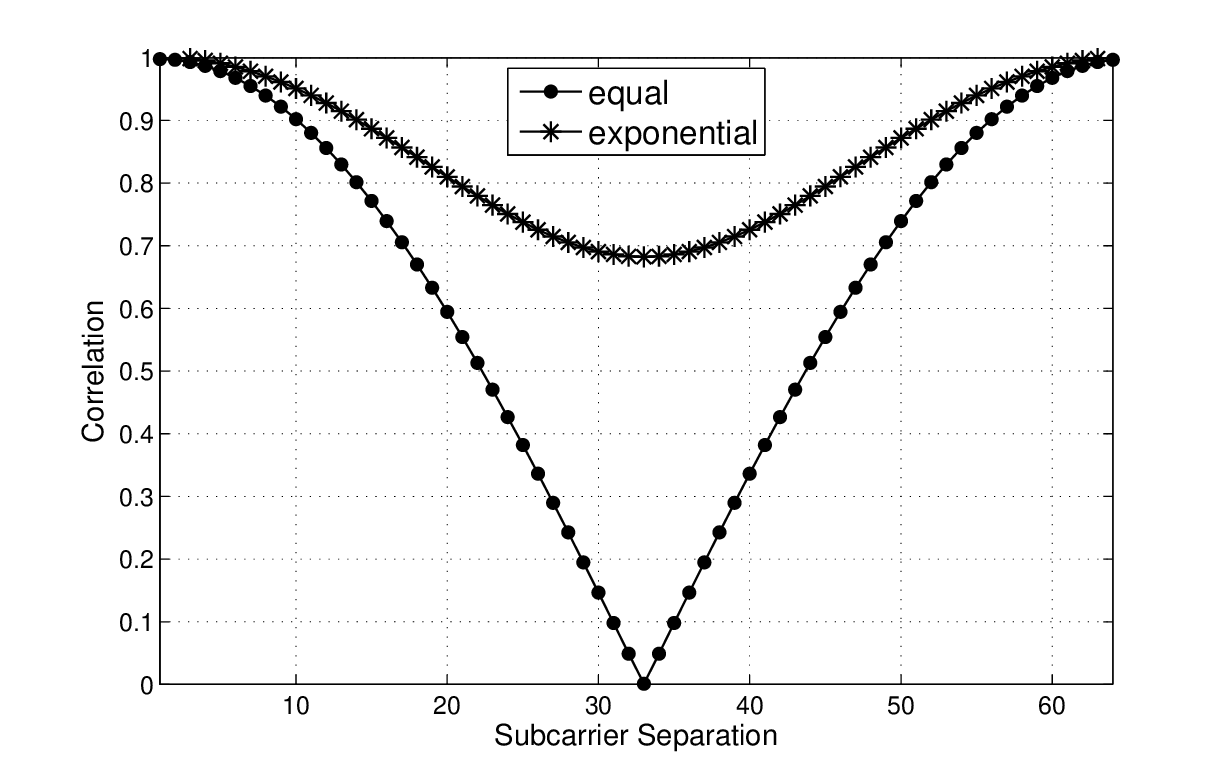}
\caption{Correlation vs. subcarrier separation for $2 \times 2$, $L=2$, $M=64$ BICMB-OFDM over equal and exponential power channel taps.}
\label{fig:correlation_2tap}
\end{figure}
\else
\begin{figure}[!t]
\centering \includegraphics[width = 1.0\linewidth]{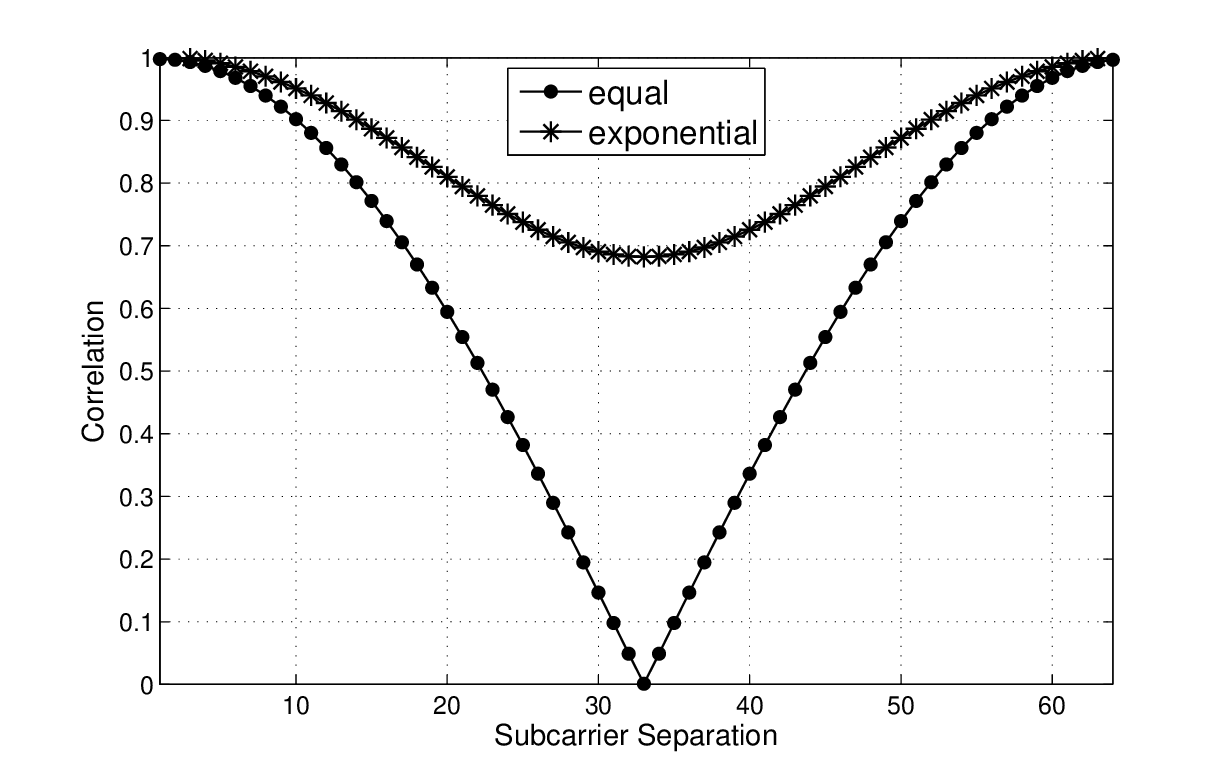}
\caption{Correlation vs. subcarrier separation for $2 \times 2$, $L=2$, $M=64$ BICMB-OFDM over equal and exponential power channel taps.}
\label{fig:correlation_2tap}
\end{figure}
\fi

Fig. \ref{fig:correlation_2tap} shows the correlation $\rho$ of two subcarriers with different separation for $2 \times 2$, $L=2$, $M=64$ BICMB-OFDM over equal and exponential power channel taps. The figure shows that the channel with exponential power taps causes stronger subcarrier correlation than equal power taps, which results in worse performance as shown in Fig. \ref{fig:exp_2tap}. 

\ifCLASSOPTIONonecolumn
\begin{figure}[!t]
\centering \includegraphics[width = 1.0\linewidth]{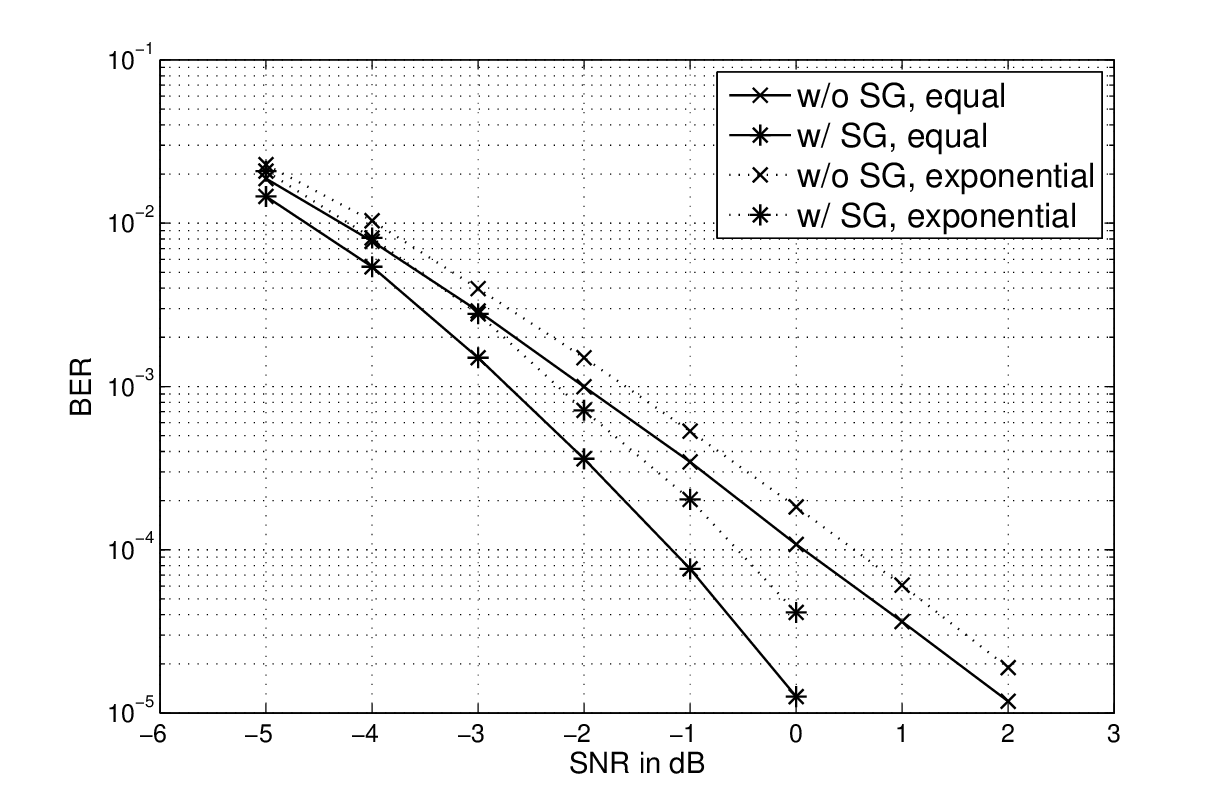}
\caption{BER vs. SNR for $2 \times 2$, $L=4$, $M=64$, $S=1$, $R_c=1/4$ BICMB-OFDM with and without subcarrier grouping over equal and exponential power channel taps.}
\label{fig:exp_4tap}
\end{figure}
\else
\begin{figure}[!t]
\centering \includegraphics[width = 1.0\linewidth]{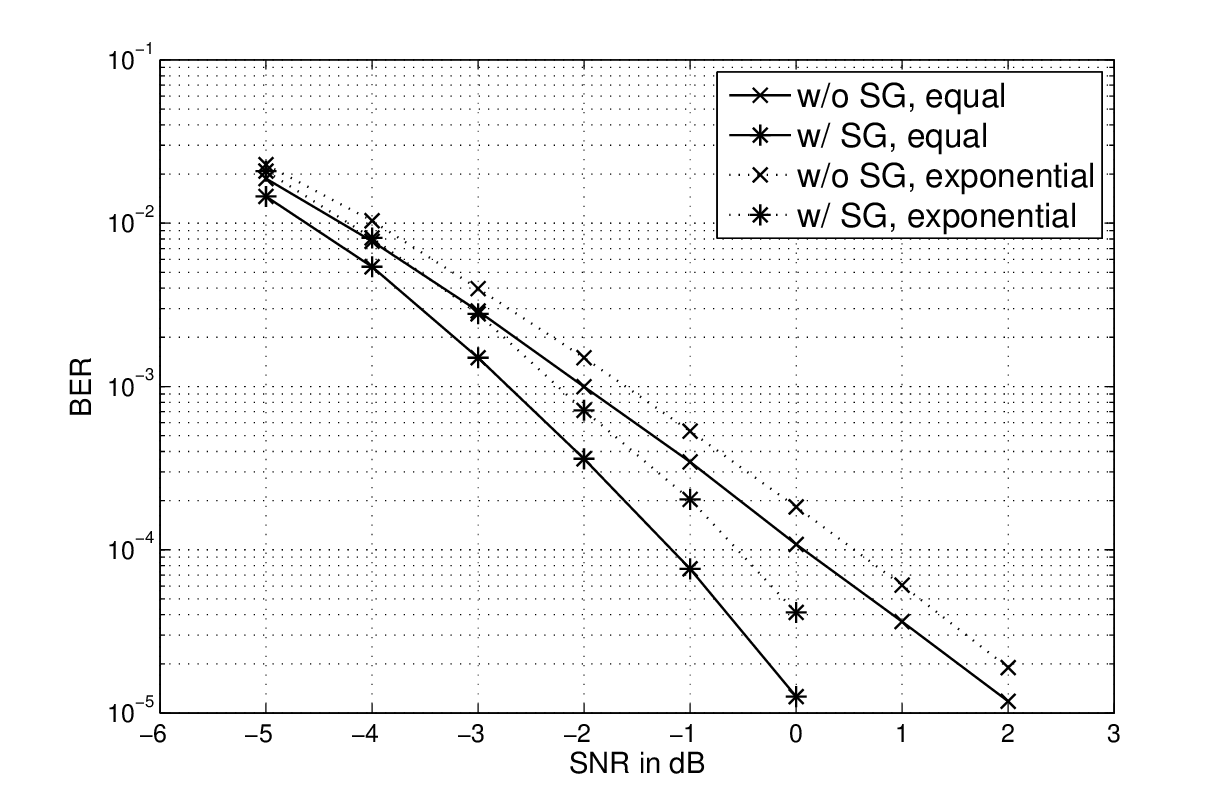}
\caption{BER vs. SNR for $2 \times 2$, $L=4$, $M=64$, $S=1$, $R_c=1/4$ BICMB-OFDM with and without subcarrier grouping over equal and exponential power channel taps.}
\label{fig:exp_4tap}
\end{figure}
\fi

Similarly to Fig. \ref{fig:exp_2tap}, Fig. \ref{fig:exp_4tap} shows the BER performance of $2 \times 2$, $L=4$, $M=64$, $S=1$, $R_c=1/4$ BICMB-OFDM with and without subcarrier grouping over equal and exponential power channel taps. The results show that the negative effect of subcarrier correlation on performance can be overcome by the subcarrier grouping technique for both equal and exponential power channel taps. For equal power channel taps, the maximum achievable diversity of $16$ is achieved by applying subcarrier grouping as discussed in Section \ref{sec:grouping} because subcarrier correlation is totally removed. On the other hand, since subcarrier correlation is not totally reduced in the case of exponential power channel taps, subcarrier grouping cannot fully restore the performance.

\ifCLASSOPTIONonecolumn
\begin{figure}[!t]
\centering \includegraphics[width = 1.0\linewidth]{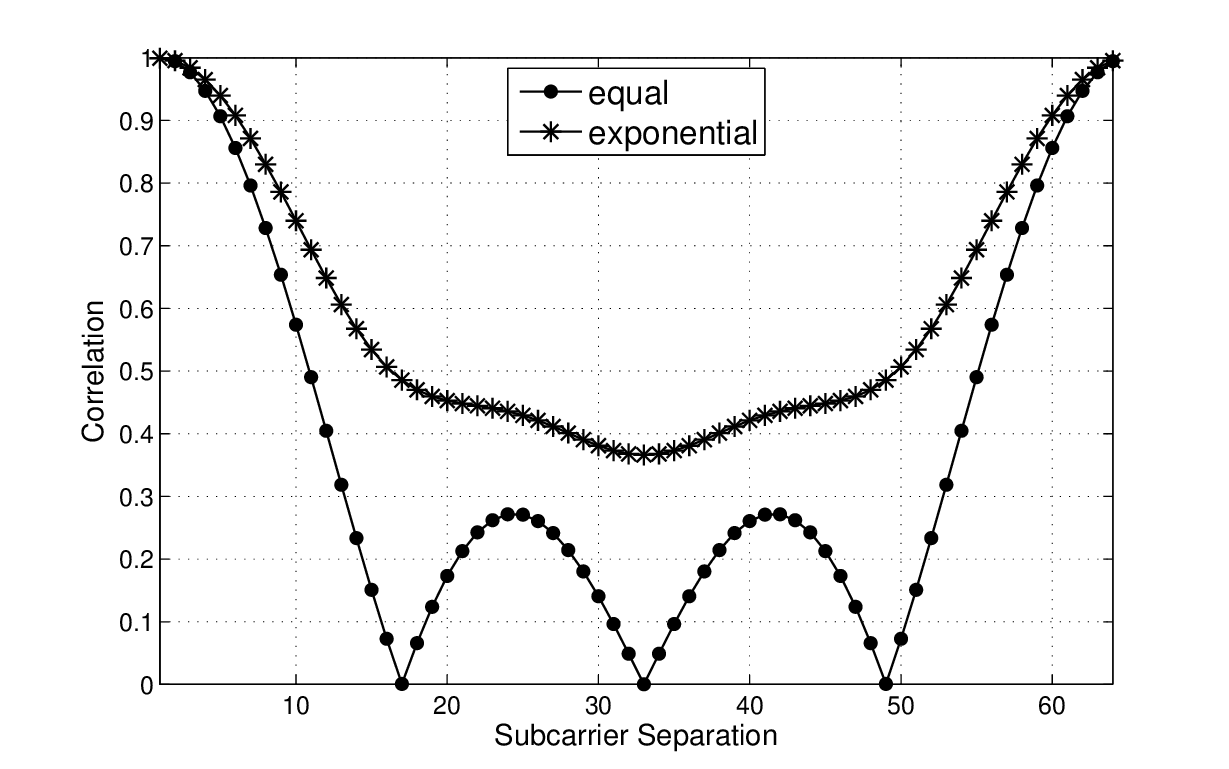}
\caption{Correlation vs. subcarrier separation for $2 \times 2$, $L=4$, $M=64$ BICMB-OFDM over equal and exponential power channel taps.}
\label{fig:correlation_4tap}
\end{figure}
\else
\begin{figure}[!t]
\centering \includegraphics[width = 1.0\linewidth]{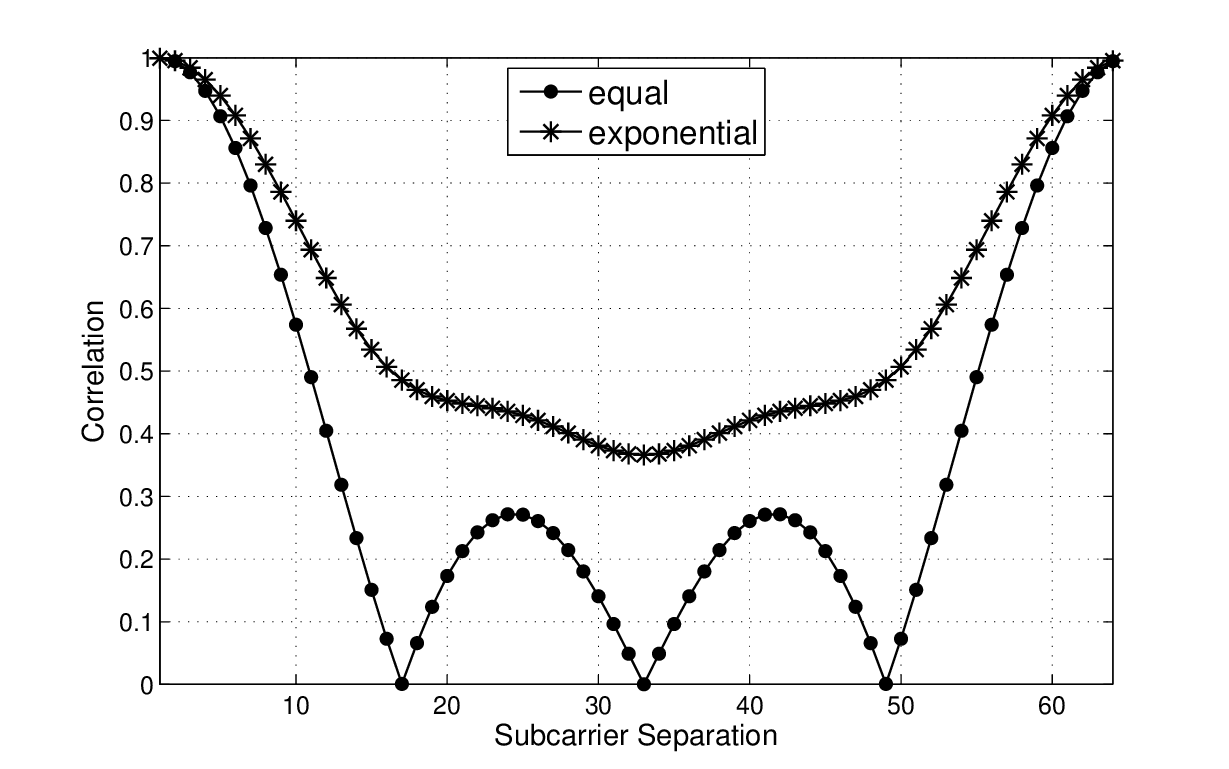}
\caption{Correlation vs. subcarrier separation for $2 \times 2$, $L=4$, $M=64$ BICMB-OFDM over equal and exponential power channel taps.}
\label{fig:correlation_4tap}
\end{figure}
\fi

Similarly to Fig. \ref{fig:correlation_2tap}, Fig. \ref{fig:correlation_4tap} shows the correlation $\rho$ of two subcarriers with different separation for $2 \times 2$, $L=4$, $M=64$ BICMB-OFDM over equal and exponential power channel taps. The figure shows that the subcarrier correlation of exponential power channel taps is larger than equal power channel taps so that it achieves worse performance as shown in Fig. \ref{fig:exp_4tap}. 

Fig. \ref{fig:expectation} verifies the negative effect of subcarrier correlation as analyzed in Section \ref{sec:correlation}, while Fig. \ref{fig:exp_2tap} and Fig. \ref{fig:exp_4tap} verify the advantage of BICMB-OFDM with subcarrier grouping over BICMB-OFDM without subcarrier grouping. Other than the subcarrier grouping technique, focus can be drawn on the design of the bit interleaver to combat the subcarrier correlation. As illustrated in Section \ref{subsec:alpha}, the bit interleaver is directly related to the $\alpha$-spectra which reflect the subcarrier distribution of the distinct bits for each error path and thereby determine the diversity. In fact, the performance degradation results from the error paths whose distinct bits are transmitted by correlated subcarriers. Therefore, negative effect of subcarrier correlation can be reduced by a properly designed bit interleaver so that the errored bits of every error event are carried on uncorrelated or weakly correlated subcarriers. However, only considering the worst-case error event is already very difficult. For BICMB-OFDM, at high SNR, the performance is dominated by the worst-case error event of the error events which have the worst diversity order. When the number of subcarriers is large, there may exist too many error events with the same worst diversity order. When subcarrier correlation exists, it is already very hard to analyze the performance of error events which involve more than two correlated subcarriers, as mentioned in Section \ref{sec:correlation}. As a result, it is even harder to determine the worst-case error event. Even if the worst-case error event can be identified, only focusing on that event is not sufficient. To lighten the negative effect of subcarrier correlation for the worst-case error event, the assigned subchannels need to be rearranged. However, the rearrangement of subchannels also affects other error events. It is probable that after the rearrangement another error event becomes the worst one, which might be even worse than the original worst case. As a result, all error events need to be considered for an interleaver design. However, because there are countless error events, it is impossible to consider them all. In fact, such an interleaver can only find a better arrangement of subcarriers to lighten the negative effect of subcarrier correlation for some error events, but the subcarrier correlation itself is not changed at all. On the other hand, the subcarrier correlation is actually reduced by subcarrier grouping, which results in better performance. As a result, the subcarrier grouping technique is apparently a better choice because it does not only achieve better performance in an easier way but also provides multi-user compatibility as explained in Section \ref{sec:grouping}. Therefore, the design of the bit interleaver to combat subcarrier correlation is not considered in this paper.

\subsection{Outdated CSI and No CSI at the Transmitter} \label{subsec:results_csi}
As presented in Section \ref{sec:system}, BICMB-OFDM requires the knowledge of CSI at the Transmitter (CSIT). However, due to the feedback delay caused by the channel-access protocols overhead or signal processing intervals, CSIT usually becomes outdated before actually being applied at the transmitter. As a result, the system performance would be significantly degraded \cite{Zhou_AMMT_CMF, Lau_CLD_UMA_OCSI, Huang_FD_MALF_TCC}. An effective approach to overcome this issue is to predict the channel at the receiver based on the past channel knowledge and thereby decide the feedback \cite{Li_MIMO_OFDM_FD, Liu_NTBS_TSFMS, Liu_STBA_MA_OFDM, Yang_TST_MIMO_LRF, Benvenuto_PCQ_BD_LF, Ramya_EB_DF_CP, Zhu_QP_PMIMO_DLF}. In particular, the performance degradation caused by outdated CSIT for multiple-antenna OFDM beamforming has been addressed with acceptable performance in \cite{Li_MIMO_OFDM_FD, Liu_STBA_MA_OFDM}. In practice, similar techniques can be applied to BICMB-OFDM. 

When CSIT is not available, space-time or space-frequency coding techniques with OFDM have been used to achieve diversity for frequency selective fading MIMO channels  \cite{Bolcskei_SFC, Su_FRFD_SFC, Shao_SFBC, Zhang_STC, Fazel_QOSTFBC}. In general, higher rate of the space-time or space-frequency code results in higher decoding complexity. Orthogonal codes introduced in \cite{Jafarkhani_STC} can achieve the same symbol-by-symbol decoding complexity as BICMB-OFDM. Other than the orthogonal codes, non-orthogonal codes increase the decoding complexity. It has been shown that with the same decoding complexity, the same rate, and the same transmit power, BICMB-OFDM can significantly outperform OFDM with orthogonal codes \cite{Akay_BICMB_OFDM ,Akay_BICMB_OFDM_CSI}, which verifies the importance of CSIT.   

In this subsection, we will first investigate the effects of outdated CSIT on BICMB-OFDM, and then we will compare the performance of BICMB-OFDM with another coded MIMO-OFDM system that does not require CSIT. All the coded MIMO-OFDM systems we will discuss have parameters of $2 \times 2$, $L=4$, $M=64$, with $4$-QAM, and have the same transmission rate and power. 

We model the outdated CSIT as in \cite{Li_SVD_TDD} which is summarized here. For the $l$th channel tap at time $t$, i.e., $\breve{\mathbf{H}}_t(l)$, the entries $\breve{h}_{u,v,t}(l)$ are assumed to be independent complex normal random variables with zero mean and variance $\sigma^2(l)$, where $u=1,\ldots, N_r$ and $v=1,\ldots, N_t$. To characterize the outdated CSIT, the channel time variation is described as $\breve{h}_{u,v,t}(l) = \rho \breve{h}_{u,v,t-\tau}(l) + \sqrt{1-\rho^2} e(l)$, where $\tau$ stands for the delay of CSIT, $\breve{h}_{u,v,t-\tau}(l)$ is the $(u,v)$th outdated channel which is known at the transmitter instead of the true channel $\breve{h}_{u,v,t}(l)$, $\rho$ denotes the time correlation coefficient, and $e(l)$ is a random variable with zero mean and variance $\sigma^2(l)$. In our simulations, $\rho$ is derived based on Jakes' model \cite{Jakes_MMC} which depends on $\tau$ and the maximum Doppler frequency shift. Specifically, we consider $\rho \approx 0.9$ for our simulations.

\ifCLASSOPTIONonecolumn
\begin{figure}[!t]
\centering \includegraphics[width = 1.0\linewidth]{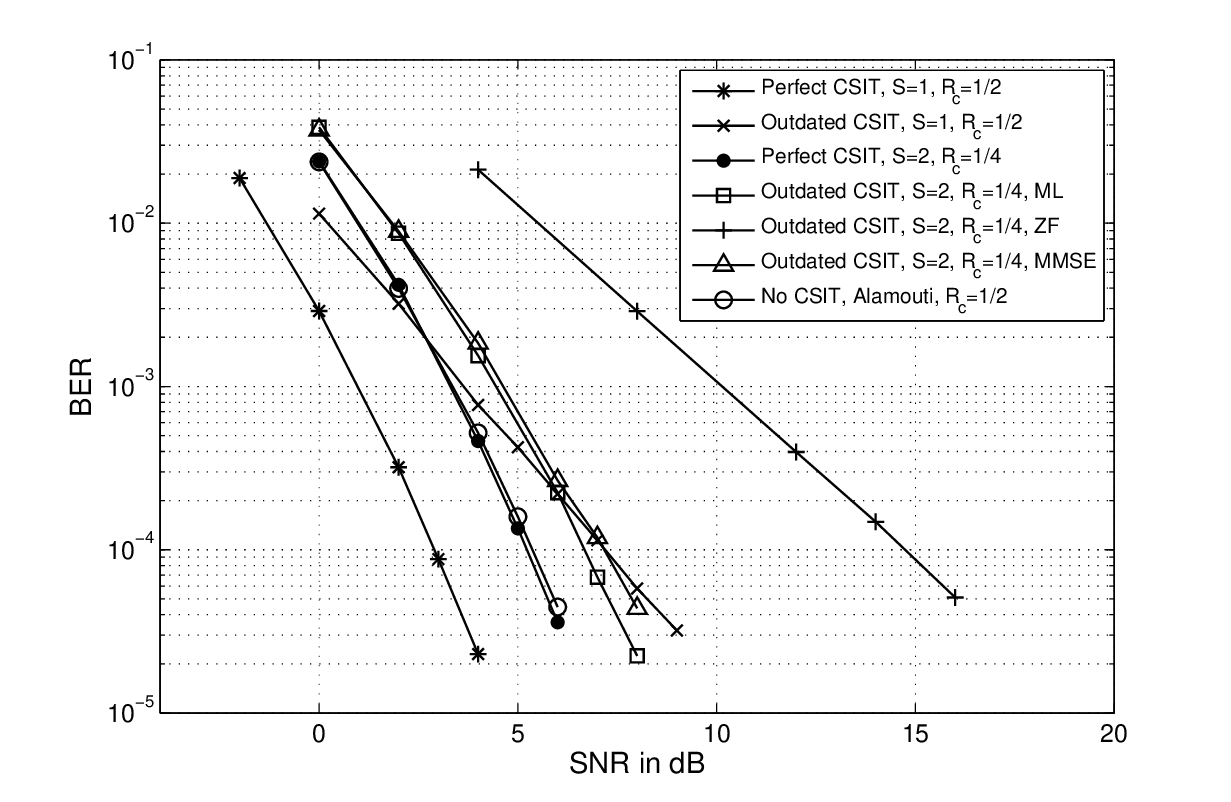}
\caption{BER vs. SNR for $2 \times 2$, $L=2$, $M=64$ coded MIMO-OFDM with perfect, outdated, and no CSIT.}
\label{fig:outdated}
\end{figure}
\else
\begin{figure}[!t]
\centering \includegraphics[width = 1.0\linewidth]{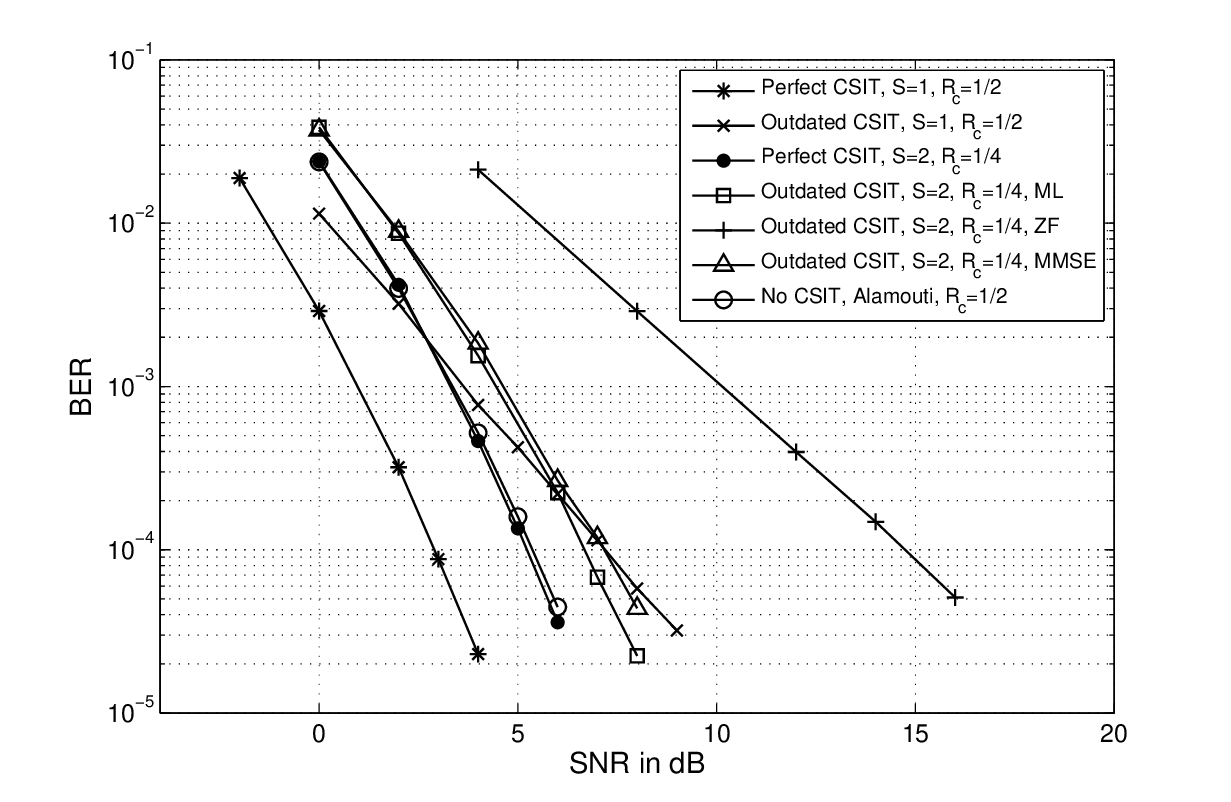}
\caption{BER vs. SNR for $2 \times 2$, $L=2$, $M=64$ coded MIMO-OFDM with perfect, outdated, and no CSIT.}
\label{fig:outdated}
\end{figure}
\fi
 
Fig. \ref{fig:outdated} shows the performance of $2 \times 2$, $L=2$, $M=64$ coded MIMO-OFDM with perfect, outdated, and no CSIT. With perfect and outdated CSIT, BICMB-OFDM introduced in this paper is employed. Two combinations of $S=1$, $R_c=1/2$, and $S=2$, $R_c=1/4$, are considered. As shown in Fig. \ref{fig:uncorrelated_2tap}, they both achieve full diversity with perfect CSIT. For $S=1$, symbol-by-symbol ML decoding is applied for both perfect and outdated CSIT. On the other hand, in the case of $S=2$, symbol-by-symbol ML decoding is only applicable for perfect CSIT. As for outdated CSIT of $S=2$, other than ML decoding which increases the decoding complexity due to joint symbol decoding from both streams, two suboptimal linear decoding techniques for coded MIMO-OFDM systems, i.e., Zero-Forcing (ZF) \cite{McKay_MIMO_BICM_ZF} and Minimum Mean Square Error (MMSE) \cite{Seethaler_MMSE_MIMO_BICM}, are also considered. When CSIT is not available, Alamouti code \cite{Jafarkhani_STC} is employed for the coded MIMO-OFDM system with $R_c=1/2$, which was shown achieving full diversity in \cite{Akay_BICMB_OFDM ,Akay_BICMB_OFDM_CSI}.  

We first compare the performance of BICMB-OFDM with perfect and outdated CSIT. In the case of $S=1$, the diversity of outdated CSIT is degraded significantly compared to perfect CSIT. Similarly, for $S=2$, the diversity of perfect CSIT is better than all three decoders with outdated CSIT. Specifically, ML with increased decoding complexity provides the best performance which can be achieved with outdated CSIT, but there is still a diversity degradation. As for suboptimal linear decoding, MMSE achieves close performance to ML while ZF suffers substantial performance loss. Note that with ML or MMSE, the sensitivity of outdated CSIT for $S=2$ is less severe than the case of $S=1$. In summary, there is a reduction of diversity with the outdated CSIT. However, systems where CSIT can be extracted readily, such as those based on Time Domain Duplexing (TDD), will not have this reduction.

We now compare the performance of BICMB-OFDM with the coded Alamouti-OFDM system which requires no CSIT. In the case of perfect CSIT, our BICMB-OFDM with $S=2$ has the same performance as the coded Alamouti-OFDM. On the other hand, our BICMB-OFDM with $S=1$ outperforms the coded Alamouti-OFDM. We note with transmit power optimization, the performance of $S=2$ BICMB-OFDM can be improved \cite{Stupia_BICMB_OFDM}, thereby can beat the coded Alamouti-OFDM. However, such a study is beyond the scope of this paper. As for outdated CSIT, BICMB-OFDM performs worse than the coded Alamouti-OFDM which requires no CSIT. In summary, when the CSIT is outdated, instead of SVD beamforming, other space-time techniques requiring no CSIT may be a better choice. On the other hand, if CSIT can be extracted readily such as in TDD, our BICMB-OFDM provides performance advantage.  


%% file: Conclusions.tex
\section{Conclusions} \label{sec:conclusions}

BICMB-OFDM combines MIMO and OFDM to achieve spatial diversity, multipath diversity, spatial multiplexing, and frequency multiplexing simultaneously for frequency selective fading MIMO channels so that it could be an important technique for broadband wireless communication. In this paper, the diversity analysis of BICMB-OFDM is carried out. As a result, the maximum achievable diversity is derived and the important $\alpha$-spectra directly determining the diversity is introduced, providing important insights of BICMB-OFDM. According to the analysis, a sufficient and necessary condition $R_cSL \leq 1$ for achieving full diversity is proved, which is very important for practical design. In addition, the negative effect of subcarrier correlation on the performance in the practical case is investigated, and subcarrier grouping is employed to overcome the performance degradation and provide multi-user compatibility. Furthermore, it is possible to employ precoding techniques so that the full diversity condition of BICMB-OFDM is not restricted to $R_cSL \leq 1$, and its diversity analysis and design criteria are available in \cite{Li_FDPD_BICMB_OFDM_J, Li_FDPD_BICMB_OFDM}.

%% file: Acknowledgment.tex
\section*{Acknowledgment}
The authors would like to thank the editor and the anonymous reviewers whose valuable comments improved the quality of the paper. 

%% file: Appendix.tex
\appendices
\section{Proof of the Smallest Degree of  $f_2(\mathbf{\Phi}_{\mathbf{p}}, \tilde{\mathbf{\Phi}}_{\tilde{\mathbf{p}}})$}
The polynomial $f_2(\mathbf{\Phi}_{\mathbf{p}}, \tilde{\mathbf{\Phi}}_{\tilde{\mathbf{p}}})$ in (\ref{eq:pdf_marginal_polynomial}) corresponds to (\ref{eq:pdf_correlated_wishart}). Because the relations $\int_{0}^{v} u^t e^{-u} \, \mathrm{d} u \leq {1 \over t+1}v^{t+1}$ and $\int_{0}^{\infty} u^t e^{-u} \, \mathrm{d} u = t!$ are valid, the smallest degree of $f_2(\mathbf{\Phi}_{\mathbf{p}}, \tilde{\mathbf{\Phi}}_{\tilde{\mathbf{p}}})$ is related to the polynomial $f_1(\mathbf{\Phi}, \tilde{\mathbf{\Phi}})$ in (\ref{eq:pdf_correlated_wishart}), which is given by (\ref{eq:pdf_correlated_wishart_polynomial}) and can be rewritten as 
\ifCLASSOPTIONtwocolumn
\begin{align}
f_1(\mathbf{\Phi}, \tilde{\mathbf{\Phi}}) & = \epsilon^{(X-Y)/2} [\prod_{u<v}^Y ( \phi_u - \phi_v ) ( \tilde{\phi}_u - \tilde{\phi}_v ) ] \nonumber \\ 
& \quad \times [ \prod_{u=1}^Y (\phi_u \tilde{\phi}_u)^{X-Y}] \mathrm{det} [\tilde{I}_{X-Y}(\epsilon \phi_u \tilde{\phi}_v)],
\label{eq:pdf_correlated_wishart_polynomial_revised}
\end{align}
\else
\begin{align}
f_1(\mathbf{\Phi}, \tilde{\mathbf{\Phi}}) = \epsilon^{(X-Y)/2} [\prod_{u<v}^Y ( \phi_u - \phi_v ) ( \tilde{\phi}_u - \tilde{\phi}_v ) ] [ \prod_{u=1}^Y (\phi_u \tilde{\phi}_u)^{X-Y}]  \mathrm{det} [\tilde{I}_{X-Y}(\epsilon \phi_u \tilde{\phi}_v)],
\label{eq:pdf_correlated_wishart_polynomial_revised}
\end{align}
\fi
where 
\begin{align}
\tilde{I}_{N}(t) = \sum_{j=0}^{\infty} {{t^j} \over j!(j+N+1)!}.
\label{eq:modified_bessel_simple}
\end{align}
Note that only the multivariate term of $f_1(\mathbf{\Phi}, \tilde{\mathbf{\Phi}})$ determining the smallest degree of  $f_2(\mathbf{\Phi}_{\mathbf{p}}, \tilde{\mathbf{\Phi}}_{\tilde{\mathbf{p}}})$ needs to be considered. The dominant term of $f_1(\mathbf{\Phi}, \tilde{\mathbf{\Phi}})$ is the one with the smallest degree and the largest eigenvalues, which depends on the dominant term of $\prod_{u<v}^Y ( \phi_u - \phi_v ) ( \tilde{\phi}_u - \tilde{\phi}_v )$ and the dominant term of $ \mathrm{det}[\tilde{I}_{X-Y}(\epsilon \phi_u \tilde{\phi}_v)]$. Obviously, the dominant term of $\prod_{u<v}^Y ( \phi_u - \phi_v ) ( \tilde{\phi}_u - \tilde{\phi}_v )$ is $ \prod_{u=1}^Y (\phi_u \tilde{\phi}_u)^{Y-u}$. On the other hand, the dominant term of $ \mathrm{det}[\tilde{I}_{X-Y}(\epsilon \phi_u \tilde{\phi}_v)]$ is $\zeta \prod_{u=1}^Y (\phi_u \tilde{\phi}_u)^{Y-u}$ where $\zeta$ is a constant, and the proof is provided in Appendix B. Therefore, the dominant term in $f_1(\mathbf{\Phi}, \tilde{\mathbf{\Phi}})$ ignoring the constant, is given by
\ifCLASSOPTIONtwocolumn
\begin{align}
\tilde{f}_1(\mathbf{\Phi}, \tilde{\mathbf{\Phi}}) = \prod_{u=1}^Y (\phi_u \tilde{\phi}_u)^{X+Y-2u}.
\label{eq:pdf_correlated_wishart_polynomial_dominant}
\end{align}
\else
\begin{align}
\tilde{f}_1(\mathbf{\Phi}, \tilde{\mathbf{\Phi}}) = \prod_{u=1}^Y (\phi_u \tilde{\phi}_u)^{X+Y-2u}.
\label{eq:pdf_correlated_wishart_polynomial_dominant}
\end{align}
\fi
Therefore, the degree of $\tilde{f}_1(\mathbf{\Phi}, \tilde{\mathbf{\Phi}})$ is 
\begin{align}
\delta_{\tilde{f}_1} = 2Y(X-1).
\label{eq:degree_original}
\end{align}
After integration of (\ref{eq:pdf_marginal}), the factor $\prod_{u=1}^{p_1-1} \phi_u^{X+Y-2u}$ and the factor $\prod_{u=1}^{\tilde{p}_1-1} \tilde{\phi}_u^{X+Y-2u}$ of $ \tilde{f}_1(\mathbf{\Phi}, \tilde{\mathbf{\Phi}})$ vanish because $\int_{0}^{\infty} u^t e^{-u} \, \mathrm{d} u = t!$. Hence,
\begin{align}
\delta_{vanished} = (p_1-1)(X+Y-p_1) + (\tilde{p}_1-1)(X+Y-\tilde{p}_1).
\label{eq:degree_vanished}
\end{align}
Meanwhile, the eigenvalues $\phi_{q_u}$ with $q_u>p_1$ and $\phi_{\tilde{q}_u}$ with $\tilde{q}_u>\tilde{p}_1$ result in increased degree because $\int_{0}^{v} u^t e^{-u} \, \mathrm{d} u \leq {1 \over t+1}v^{t+1}$. Therefore,
\begin{align}
\delta_{added} = 2Y-W-\tilde{W}-p_1-\tilde{p}_1+2.
\label{eq:degree_increased}
\end{align}
As a result, the smallest degree of $f_2(\mathbf{\Phi}_{\mathbf{p}}, \tilde{\mathbf{\Phi}}_{\tilde{\mathbf{p}}})$ is
\ifCLASSOPTIONtwocolumn
\begin{align}
\delta & = \delta_{\tilde{f}_1} - \delta_{vanished} + \delta_{added} \nonumber \\
& = (X-p_1+1)(Y-p_1+1) - W \nonumber \\
& \quad +(X-\tilde{p}_1+1)(Y-\tilde{p}_1+1) - \tilde{W}.
\label{eq:degree}
\end{align}
\else
\begin{align}
\delta & = \delta_{\tilde{f}_1} - \delta_{vanished} + \delta_{added} \nonumber \\
&= (X-p_1+1)(Y-p_1+1)+(X-\tilde{p}_1+1)(Y-\tilde{p}_1+1)-W-\tilde{W}.
\label{eq:degree}
\end{align}
\fi

\section{proof of the Dominant term of $\mathrm{det}[\tilde{I}_{X-Y}(\epsilon \phi_u \tilde{\phi}_v)]$}
When $Y=1$,
\begin{align}
\mathrm{det}[\tilde{I}_{X-Y}(\epsilon \phi_u \tilde{\phi}_v)] = \sum_{j=0}^{\infty} {{(\epsilon \phi_1 \tilde{\phi}_1)^j} \over j!(j+X)!}
\label{eq:dominant_1}
\end{align}
and the dominant term is $1/X!$.

When $Y=2$,  
\ifCLASSOPTIONtwocolumn
\begin{align}
\mathrm{det}[\tilde{I}_{X-Y}(\epsilon \phi_u \tilde{\phi}_v)] & = \tilde{I}_{X-Y}(\epsilon \phi_1 \tilde{\phi}_1) \tilde{I}_{X-Y}(\epsilon \phi_2 \tilde{\phi}_2) \nonumber \\ 
& \quad - \tilde{I}_{X-Y}(\epsilon \phi_1 \tilde{\phi}_2) \tilde{I}_{X-Y}(\epsilon \phi_2 \tilde{\phi}_1) \nonumber \\
& = \sum_{u=1}^{2} \sum_{v=1}^{2} (-1)^{u+v} \nonumber \\
& \left[ \sum_{j=0}^{\infty} \sum_{k>j}^{\infty}{{(\epsilon \phi_u \tilde{\phi}_v)^j} \over j!(j+X-1)!} {{(\epsilon \phi_{3-u} \tilde{\phi}_{3-v})^k} \over k!(k+X-1)!} \right]
\label{eq:dominant_2} 
\end{align}
\else
\begin{align}
\mathrm{det}[\tilde{I}_{X-Y}(\epsilon \phi_u \tilde{\phi}_v)] & = \tilde{I}_{X-Y}(\epsilon \phi_1 \tilde{\phi}_1) \tilde{I}_{X-Y}(\epsilon \phi_2 \tilde{\phi}_2) - \tilde{I}_{X-Y}(\epsilon \phi_1 \tilde{\phi}_2) \tilde{I}_{X-Y}(\epsilon \phi_2 \tilde{\phi}_1) \nonumber \\
& = \sum_{u=1}^{2} \sum_{v=1}^{2} (-1)^{u+v} \left[ \sum_{j=0}^{\infty} \sum_{k>j}^{\infty}{{(\epsilon \phi_u \tilde{\phi}_v)^j} \over j!(j+X-1)!} {{(\epsilon \phi_{3-u} \tilde{\phi}_{3-v})^k} \over k!(k+X-1)!} \right]
\label{eq:dominant_2}
\end{align}
\fi
and the dominant term is $\epsilon \phi_1 \tilde{\phi}_1 / \left[X!(X-1)!\right]$.

When $Y \geq 3$,
\ifCLASSOPTIONtwocolumn
\begin{align}
\mathrm{det}[\tilde{I}_{X-Y}(\epsilon \phi_u \tilde{\phi}_v)] & = \sum_{u=1}^{Y} \sum_{v=1}^{Y} (-1)^{u+v} \nonumber \\
& \left[ \prod_{k=1}^{Y} \sum_{j_k=0}^{\infty} {(\epsilon \phi_{u_k} \tilde{\phi}_{v_k})^{j_k} \over {j_k}!(j_k+X-Y+1)!} \right]_{j_k<j_{k+1}}
\label{eq:dominant_general}
\end{align}
\else
\begin{align}
\mathrm{det}[\tilde{I}_{X-Y}(\epsilon \phi_u \tilde{\phi}_v)] = \sum_{u=1}^{Y} \sum_{v=1}^{Y} (-1)^{u+v} \left[ \prod_{k=1}^{Y} \sum_{j_k=0}^{\infty} {(\epsilon \phi_{u_k} \tilde{\phi}_{v_k})^{j_k} \over {j_k}!(j_k+X-Y+1)!} \right]_{j_k<j_{k+1}}
\label{eq:dominant_general}
\end{align}
\fi
where $u_k=[(u+k-2) \bmod Y]+1$ and $v_k=[(v+k-2) \bmod Y]+1$, and the dominant term is $\zeta \prod_{k=1}^Y (\phi_k \tilde{\phi}_k)^{Y-k}$ with $\zeta =  \prod_{k=1}^Y \epsilon^{Y-k} / \left[(Y-k)!(X-k+1)! \right]$.
